\documentclass[aps,prb,twocolumn,showpacs,superscriptaddress,10pt]{revtex4-1}
\usepackage{soul}
\usepackage[english]{babel}
\usepackage{amsmath,amsfonts,amsthm,braket}
\usepackage{morefloats}
\usepackage{bm}
\usepackage{graphicx,color}
\usepackage[outdir=./]{epstopdf}

\newcommand*\chem[1]{\ensuremath{\mathrm{#1}}}
\newcommand*\ham{\hat{H}}
\newcommand*\crt[1]{\hat{a}^\dagger_{#1}}
\newcommand*\dst[1]{\hat{a}^{\phantom{\dagger}}_{#1}}
\newcommand*\vett[1]{{\bf{#1}}}

\newtheorem{Theorem}{Theorem}

\usepackage{hyperref}
\hypersetup{breaklinks=true}
\begin{document}

\author{Mario Motta}
\affiliation{Division of Chemistry and Chemical Engineering, California Institute of Technology, Pasadena, CA 91125, USA}

\author{Shiwei Zhang}
\altaffiliation{shiwei@wm.edu}
\affiliation{Department of Physics, the College of William and Mary, Williamsburg, VA 23187, USA}

\title{Ab initio computations of molecular systems by the auxiliary-field quantum Monte Carlo method}
\date{\today}

\begin{abstract}
The auxiliary-field quantum Monte Carlo (AFQMC) method provides a computational framework
for solving the time-independent Schr\"odinger equation in atoms, molecules, solids,
and a variety of model systems.
AFQMC has recently witnessed remarkable growth, especially as a tool for electronic structure
computations in real materials.
The method has demonstrated excellent accuracy across a variety of correlated electron systems.
Taking the form of stochastic evolution in a manifold of non-orthogonal Slater determinants,
the method resembles an ensemble of density-functional theory (DFT) calculations in the presence of
fluctuating external potentials.
Its computational cost scales as a low-power of system size, similar to the corresponding
independent-electron calculations.
Highly efficient and intrinsically parallel, AFQMC is able to take full advantage of contemporary
high-performance computing platforms and numerical libraries.
In this review, we provide a self-contained introduction to the exact and constrained variants of
AFQMC, with emphasis on its applications to the electronic structure in molecular
systems.
Representative results are presented, and theoretical foundations and implementation details
of the method are discussed.
\end{abstract}

\maketitle

\section{Introduction}

A central challenge 
in the fields of condensed matter physics, quantum chemistry, and materials science
is to determine the quantum-mechanical behavior of many interacting 
electrons and nuclei. Often relativistic effects and the coupling between the dynamics 
of electrons and nuclei can be neglected, or treated separately. 
Within this approximation, the many-electron wave function can be found by solving the 
time-independent Schr\"odinger equation for the Born-Oppenheimer Hamiltonian
\cite{Born_1927,Szabo_book_1989}
\begin{equation}
\label{eq:bo_ham}
\begin{split}
\ham = &-\frac{1}{2} \, \sum_{i=1}^N \frac{\partial^2}{\partial {\bf{r}}_i^2} + 
\sum_{i<j=1}^N \frac{1}{|{\bf{r}}_i - {\bf{r}}_j|} 
            - \sum_{i=1}^N \sum_{a=1}^{N_n} \frac{Z_a}{|{\bf{r}}_i - {\bf{R}}_a|} + \\
            &+ \sum_{a<b=1}^{N_n} \frac{Z_a Z_b}{|{\bf{R}}_a - {\bf{R}}_b|} \,,
\end{split}
\end{equation}
where ${\bf{r}}_i$ is the position of electron $i$, and $\vett{R}_a$ the position of nucleus 
$a$, with charge $Z_a$. The numbers of electrons and nuclei are $N$ 
and $N_n$, respectively.
The nuclear positions $\vett{R}_1 \dots \vett{R}_{N_n}$ are held fixed.
We use atomic units throughout.
The main obstacle to the investigation of chemical systems is that, in general, the computational 
cost of finding the exact ground state of Eq.~\eqref{eq:bo_ham} grows combinatorially with the size of 
the studied system \cite{Troyer_PRL_2004,Schuch_NAT_2009}.
This limitation has so far precluded exact studies for all but small molecular systems and 
motivated the development of approximate methods.

By far the most widely used approximate methods are independent-electron 
approaches, based on the celebrated density-functional theory (DFT) 
\cite{Martin_book_2004,Kohn_RMP_1999}. 
These are standard tools for electronic structure calculations in diverse areas 
across multiple disciplines, with sophisticated computer software packages available. 
The success of DFT-based approaches has been exceptional.
Their difficulties  are also well-documented, especially in so-called strongly correlated 
systems, for example in many transition-metal oxides. 

For molecular systems, a hierarchy of quantum chemistry (QC) methods have been 
developed, which allow systematic improvement in accuracy, at increasing computational cost. 
For example, one of the most accurate QC methods is coupled-cluster CCSD(T) 
\cite{Paldus_ACP_1999, Bartlett_RMP_2007, Shavitt_book_2009}, 
scaling as the seventh power of the system size.
There are several very promising new approaches, including density-matrix
renormalization group (DMRG)
\cite{White_PRL_1992,White_JCP_1999,Chan_JCP_2002,Olivares_JCP_2015,Chan_JCP_2016}
and full-configuration interaction quantum Monte Carlo (FCIQMC) 
\cite{Booth_JCP_2009,Booth_NAT_2013}, which have dramatically expanded the size
of the molecular systems that can be treated with very high accuracy.
However the computational cost of these methods, in the
most general case, still scales exponentially with system size. 
A recent benchmark study has been performed on the linear hydrogen chain 
\cite{Motta_PRX_2017}, including many QC methods, diffusion Monte Carlo
\cite{Reynolds_JCP_1982,Foulkes_RMP_2001}, 
and auxiliary-field quantum Monte Carlo (AFQMC).

As a general electronic structure approach, the AFQMC method \cite{Zhang_PRL90_2003}
has several unique and attractive features. 
We hope that this will become evident following the description of the method below.
As a brief overview, AFQMC combines stochastic sampling with the machinery of DFT.
It is a non-perturbative approach that balances accuracy and computational scaling. 
Utilizing a field-theoretic framework and casting the projection of the many-body ground 
state or the finite-temperature partition function as a random walk in non-orthogonal 
Slater determinant space, AFQMC takes the form of a linear combination of DFT calculations 
in fluctuating auxiliary fields. 
It can incorporate many techniques used by independent-particle methods such as DFT 
or Hartree-Fock (HF). 
AFQMC is naturally parallelizable and very well suited for large-scale computing platforms. 
Though more demanding than DFT or HF, AFQMC has computational cost growing as the third
(in some cases fourth) power of the system size, enabling applications to large systems. 

The review is structured as follows. In the next section we discuss the AFQMC method in detail:
the key technical backgrounds are described in several subsections; this is followed by 
discussions of the sign and phase problem, and of the phaseless AFQMC approach which 
removes the phase problem in electronic systems; a step-by-step algorithmic outline
is then provided. Then in  the next section, we describe several recent 
methodological advances with AFQMC, including frozen-core, downfolding and embedding 
approaches, back-propagation to compute observables and correlation functions in electronic 
systems, and correlated sampling to improve efficiency in computing energy differences. 
Finally we briefly discuss the outlook of AFQMC as an electronic structure method, 
the advantages it brings and some directions for future development.
Additional mathematical background and algorithmic details are included in the Appendix.

\section{The AFQMC method}

Using the formalism of second quantization, the Hamiltonian \eqref{eq:bo_ham} can be written as
\begin{equation}
\begin{split}
\label{eq:bo_ham2}
\ham &= H_0 + \ham_1 + \ham_2 
         = \\
         &= H_0 + \sum_{ \substack{pq \\ \sigma} } h_{pq} \, \crt{p\sigma} \dst{q\sigma} + 
         \frac{1}{2} \sum_{ \substack{pqrs \\ \sigma\tau} } v_{pqrs} \, \crt{p\sigma} \crt{q\tau} \dst{s\tau} \dst{r\sigma} \, ,
\end{split}
\end{equation}
where $H_0 = \sum_{a<b=1}^{N_n} \frac{Z_a Z_b}{|{\bf{R}}_a - {\bf{R}}_b|}$ is the inter-nuclear 
repulsion, 
$\crt{p\sigma},\dst{q\tau}$ are fermionic creation and annihilation operators associated to an 
orthonormal 
basis $\left\{ \varphi_{p} \right\} _{p=1}^{M}$ of one-electron molecular orbitals (MOs).
These basis functions are typically real (although complex basis functions can be 
handled straightforwardly in AFQMC), and 
\begin{equation}
\label{eq:bo_ham_matel}
\begin{split}
h_{pq} &= \int d{\bf r} \, \varphi_p({\bf r})\left(- \frac{1}{2} \frac{\partial^2}{\partial {\bf{r}}^2 }-
\sum_{a=1}^{N_n} \frac{Z_{a}}{|{\bf r}-{\bf R}_{a}|} \right)\varphi_q^{\,}({\bf r})\,,  \quad  \\
v_{pqrs} &=
\int d{\bf r} d{\bf r}^{\prime} \varphi_p({\bf r})\varphi_q\left({\bf r}^{\prime}\right) \,
\frac{ 1 }{|{\bf r}-{\bf r}^{\prime}|} \, \varphi_r^{\,}({\bf r})\varphi_s^{\,}\left({\bf r}^{\prime}\right)\,, \quad  \\
\end{split}
\end{equation}
are the matrix elements of the one- and two-body parts of \eqref{eq:bo_ham2} in that basis. 
$v_{pqrs}$ is often denoted as $( p r  | q s )$ in chemistry, and  $\langle p q | r s \rangle$  in physics.

Most applications of AFQMC in molecular systems so far have relied on basis sets of 
MOs obtained by orthonormalizing a set $\left\{ \varphi^\prime_{i} \right\} _{i=1}^M$ 
of atom-centered atomic orbitals (AOs) expressed by standard Gaussian basis functions
\cite{Szabo_book_1989}.
The overlaps $S^\prime_{pq}= \int d{\bf r}\,\varphi^\prime_p({\bf r}) \varphi^\prime_q {\,}({\bf r})$ 
and matrix elements $h^\prime_{pq}$, $v^\prime_{pqrs}$ in the AO basis are computed and 
output by standard quantum chemistry softwares
\cite{Schmidt_JCC14_1993,Valiev_CPC181_2010,Sun_WIRES_2017}. 
The matrix elements of Eq.~\eqref{eq:bo_ham_matel} result from a straightforward change of basis.

The use of a finite basis set unavoidably introduces an approximation, which can be 
removed by extrapolating results to the complete basis set (CBS) limit. This is common 
to all quantum chemistry methods. For Gaussian bases, it is necessary to perform 
calculations with increasingly large basis sets designed to allow systematic convergence 
to the CBS limit. AFQMC scales as $M^3$ to $M^4$, as further discussed below. 
Compared to high-level quantum chemistry methods such as CCSD(T), this is 
advantageous for reaching the CBS. We typically use well-established bases 
(e.g. cc-pVxZ for light atoms \cite{Dunning_JCP90_1989,Woon_JCP99_1993},  or
cc-pVnZ-PP \cite{Peterson_JCP119_2003,Peterson_JCP119_2003_b}) and empirical 
extrapolation formulas \cite{Feller_JCP96_1992,Helgaker_JCP106_1997} for CBS extrapolations.

AFQMC allows for the choice of any one-electron basis suitable for the  problem.
In addition to Gaussians, 
AFQMC has also been implemented using plane-wave and pseudopotentials
\cite{Suewattana_PRB75_2007,Zhang_CPC169_2005,Kwee_PRL100_2008,Purwanto_PRB80_2009,
Ma_PRL114_2015,Ma_PRB_2016}, more suited for calculations in solids. 
As we discuss below, one could also use Kohn-Sham orbitals derived from a plane-wave 
DFT calculation as basis sets, which provides a kind of hybrid between the plane-wave 
and the MO approach.

\subsection{Properties of (non-orthogonal) Slater determinants}

We first review some properties of Slater determinants (SDs) important to the formulation 
and implementation of AFQMC. A single SD is written as
\begin{equation}
\label{eq:SDdef}
|\Psi\rangle
= 
\prod_{i=1}^{N_\uparrow}     \crt{u_i \uparrow} 
\prod_{i=1}^{N_\downarrow} \crt{v_i \downarrow} |\emptyset\rangle\,,
\end{equation}
for $N_\uparrow$ and $N_\downarrow$ particles with spin up  and down respectively, 
occupying the orbitals
$|u_{i} \rangle =\sum_{p} \left(U_\uparrow\right)_{pi}      \, |\varphi_{p} \rangle$ 
and 
$|v_{i} \rangle =\sum_{p}  \left(U_\downarrow\right)_{pi} \, |\varphi_{p} \rangle$.
SDs are exact eigenfunctions of one-body Hamiltonians, including those of non-interacting 
systems and independent-electron Hamiltonians from DFT or mean-field HF.

There exists a key distinction between the SDs in AFQMC and in standard quantum chemistry methods. 
In quantum chemistry, SDs involved are orthogonal, as they are built from a common set of 
reference orbitals (e.g. occupied and virtual HF orbitals) through different excitations of the 
electrons. 
In AFQMC, SDs are non-orthogonal to each other, as the occupied orbitals $U_\sigma$ 
of each SD are continuously changed (via rotations of the occupied orbitals) during the random walk. 
Some properties of non-orthogonal SDs are summarized below. 
(Additional details can be found in the Appendix.)
\begin{Theorem}
\label{thm:slater}
The following properties hold for any two non-orthogonal SDs $\Phi$ and $\Psi$, parametrized by 
the matrices $V_\sigma$ and $U_\sigma$, respectively, and for any spin-independent one-particle 
operator 
$\hat{A} = \sum_\sigma \sum_{pq} A^\sigma_{pq} \crt{p\sigma} \dst{q \sigma}$

\begin{enumerate}
\item Overlap: $\langle \Phi | \Psi \rangle 
= \prod_{\sigma} \, \mathrm{det}\left( V^\dagger_\sigma \, U_\sigma \right)$\,.
\item One-particle reduced density matrix:
\begin{equation}
\label{thm:1rdm}
G^{\Phi\Psi}_{p\sigma, q\tau} \equiv 
\frac
{ \langle \Phi | \crt{p\sigma} \dst{q\tau} | \Psi \rangle }
{ \langle \Phi | \Psi \rangle } = 
\delta_{\sigma\tau} \, \left( U_\sigma (V_\sigma^\dagger U_\sigma)^{-1} V_\sigma^\dagger \right)_{qp}\,.
\end{equation}
\item Generalized Wick's theorem \cite{Wick_PR80_1950,Balian_NC64_1969}:
\begin{equation}
\label{thm:2rdm}
\frac
{ \langle \Phi | \crt{p\sigma} \crt{q\tau} \dst{s\tau} \dst{r\sigma} | \Psi \rangle }
{ \langle \Phi | \Psi \rangle } 
= 
G^{\Phi\Psi}_{p\sigma,r\sigma} G^{\Phi\Psi}_{q\tau,s\tau} - 
G^{\Phi\Psi}_{p\sigma,s\tau}     G^{\Phi\Psi}_{q\tau,r\sigma}\,.
\end{equation}
\item Thouless theorem \cite{Thouless_NP21_1960,Thouless_NP22_1961}: 
the state $e^{\hat{A}}|\Psi\rangle = |\Psi^\prime \rangle$ is a SD parametrized by the matrices
\begin{equation}
\label{thm:thouless}
\left (U^\prime_\uparrow \right) = e^{A^\uparrow} U_\uparrow\,,
\quad \quad
\left (U^\prime_\downarrow \right) = e^{A^\downarrow} U_ \downarrow\,.
\end{equation}
\item Properties 2 and 3 remain invariant under orthonormalization of an SD, $|\Psi\rangle
\rightarrow | \Psi^\prime \rangle$,
where the orbitals in  $|\Psi^\prime\rangle$ are obtained by orthonormalizing those in $|\Psi\rangle$ 
(e.g., by a modified Gram-Schmidt procedure).

\end{enumerate}
\end{Theorem}

\subsection{Ground-state projection}

The phaseless AFQMC method is a projective quantum Monte Carlo (QMC) method. Projective methods 
are based on the observation that the ground state $\Phi_{0}$ of a Hamiltonian $\ham$ is the 
asymptotic solution of the imaginary-time Schr\"odinger equation
\begin{equation}
\label{eq:imt}
-\partial_{\beta}|\Psi(\beta)\rangle
=
\left(\ham-\mathcal{E}_{0}\right)|\Psi(\beta)\rangle\,,\quad\quad\Psi(0)
=
\Psi_{I}\,,
\end{equation}
where $\Psi_{I}$ is any initial wavefunction
non-orthogonal to $\Phi_{0}$, and $\mathcal{E}_{0}$ is the ground-state energy of $\ham$. 
The formal solution of \eqref{eq:imt} is obtained applying the imaginary-time propagator 
$e^{-\beta\left(\ham-\mathcal{E}_{0}\right)}$ to $\Psi_{I}$,
\begin{equation}
\label{eq:togs}
\begin{split}
|\Psi(\beta)\rangle=e^{-\beta\left(\ham-\mathcal{E}_{0}\right)}|\Psi_{I}\rangle
\xrightarrow[\beta\to \infty]{}  |\Phi_{0}\rangle\langle\Phi_{0}|\Psi_{I}\rangle\,.
\end{split}
\end{equation}
The ground-state energy $\mathcal{E}_{0}$ is unknown but it can be estimated adaptively 
as the calculation progresses, for example with the growth estimator discussed in the Appendix.

\begin{figure*}[ht!]
\begin{center}
\includegraphics[width=0.8\textwidth]{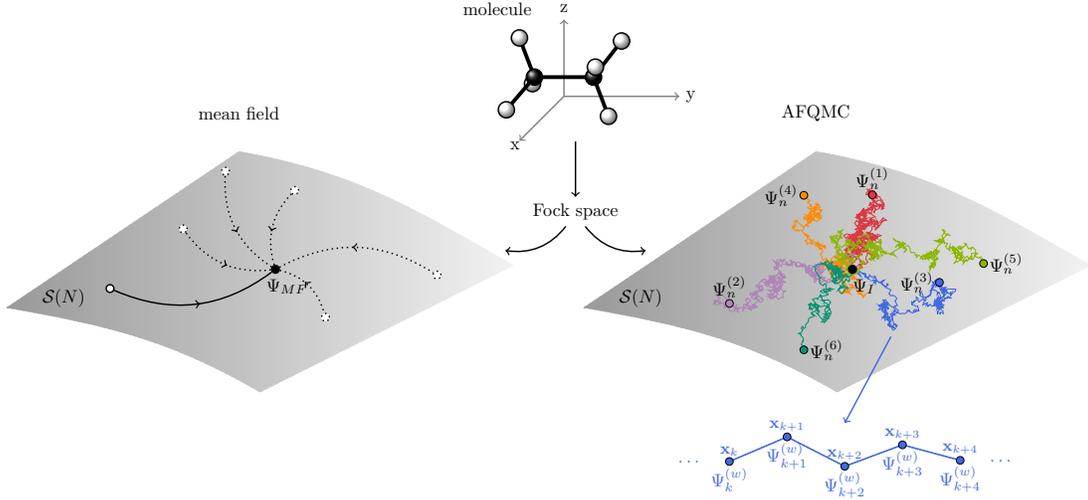}
\caption{(color online) Pictorial illustration of the AFQMC algorithm, and of its beyond-mean-field nature.
Left: independent-electron methods (e.g., HF or DFT) provide an approximation of the ground 
state wave function through a deterministic evolution in a manifold (gray surface) of Slater determinants 
$\mathcal{S}(N)$ (black curves converging to $\Psi_{MF}$).
Right: AFQMC represents the ground state as a stochastic linear combination of Slater determinants by
mapping the electron-electron interaction onto a fluctuating external potential, and the imaginary-time
evolution onto an open-ended random walk in $\mathcal{S}(N)$ (colored curves departing from $\Psi_I$).}
\label{fig:basic}
\end{center}
\end{figure*}

In general it is often as difficult to realize the imaginary-time propagation in Eq.~\eqref{eq:togs}
as it is to solve the time-independent Schr\"odinger equation.
However, for an independent-electron method such as HF, where $\ham$ is replaced by
\begin{equation}
\ham_{\Psi}=H_{0}+\ham_{1}+
\frac{1}{2} \sum_{ \substack{pqrs \\ \sigma\tau} } v_{pqrs} 
\left( \crt{p\sigma} \dst{r\sigma} G^{\Psi\Psi}_{q\tau,s\tau} - 
\crt{p\sigma} \dst{s\tau} G^{\Psi\Psi}_{q\tau,r\sigma} \right)\,,
\end{equation}
Eq.~\eqref{thm:thouless} guarantees that the imaginary-time projection turns into the motion 
of a single SD along the manifold $\mathcal{S}(N)$ of Slater determinants, 
as illustrated in Fig.~\ref{fig:basic}.
(Of course a special case would be the non-interacting system, when the 2-body part of the 
Hamiltonian vanishes.) Similarly, a DFT calculation can be cast in this way, with the projection 
in imaginary-time as an alternative for the non-linear minimization of the energy 
$\langle \Psi | \hat{H}_{\rm DFT}(\Psi) | \Psi \rangle$. 

AFQMC maps the imaginary-time projection of a many-body Hamiltonian onto a stochastic process 
in the manifold of Slater determinants, as sketched in Fig.~\ref{fig:basic}. 
The procedure to realize this mapping is outlined below.
There are different ways to view the mapping.
It can be seen as a generalization of the aforementioned single SD projection in DFT or HF, where 
many-body correlation effects is recovered by random sampling around a typical ``mean-field''.
Alternatively, the mapping can be seen as a different and complementary approach to fixed-node 
diffusion Monte Carlo
\cite{Reynolds_JCP77_1982,Foulkes_RMP73_2001}, which simulates the imaginary-time projection 
as a random walk in electron coordinate space, or Green's function Monte Carlo 
\cite{Trivedi_PRB_1989,Buonaura_PRB_1998} 
and FCIQMC \cite{Booth_JCP_2009,Booth_NAT_2013}, 
which carry out a similar projection [using $(c-\hat{H})$] in fermion configuration space 
(i.e., orthogonal SDs).
As we show below, the use of overcomplete, non-orthogonal  SDs in AFQMC significantly 
reduces the severity of the QMC fermion sign problem 
\cite{Feynman_book_1965,Loh_NATO41_1990,Schmidt_book_1984} in many situations.

\subsubsection{The Hubbard-Stratonovich transformation}

The first step to provide a stochastic representation of the imaginary-time projection is often 
a decomposition in short-time propagators,
\begin{equation}
e^{-\beta\left(\ham-\mathcal{E}_{0}\right)}
=
\left(e^{-\Delta\tau\left(\ham-\mathcal{E}_{0}\right)}\right)^{n}\,,\quad\quad\Delta\tau
=
\frac{\beta}{n}\,.\quad 
\end{equation}
Within AFQMC, one first expresses the Hamiltonian as
\begin{equation}
\label{eq:hamsum}
\ham = H_{0} + \hat{v}_{0} - \frac{1}{2} \sum_{\gamma=1}^{N_\gamma} \hat{v}_{\gamma}^{2}\,, \,\, 
\end{equation}
with $\hat{v}_{0}$, $\hat{v}_{\gamma}$ one-body operators, explicitly given in the next Section.
The short-time propagator is then approximated by the use of a Trotter breakup 
\cite{Trotter_AMS10_1959,Suzuki_PTP56_1976}
\begin{equation}
\label{eq:primitive}
\resizebox{\columnwidth}{!}{%
$e^{-\Delta\tau\left(\hat{H}-\mathcal{E}_{0}\right)} = e^{-\Delta\tau(H_{0}-\mathcal{E}_{0})} e^{-\frac{\Delta\tau}{2}\hat{v}_{0}} e^{\frac{\Delta\tau}{2}\sum_{\gamma}\hat{v}_{\gamma}^{2}} e^{-\frac{\Delta\tau}{2}\hat{v}_{0}}+\mathcal{O}\left(\Delta\tau^{3}\right)$%
}
\end{equation}
and the Hubbard-Stratonovich transformation \cite{Hubbard_PRL3_1959,Stratonovich_SPD2_1958} 
(see Appendix),
\begin{equation}
\label{eq:hs}
e^{\frac{\Delta\tau}{2}\sum_{\gamma} \hat{v}_{\gamma}^{2}} 
= 
\prod_{\gamma}
\int\frac{dx_{\gamma}}{\sqrt{2\pi}}e^{-\frac{x_{\gamma}^{2}}{2}}
e^{\sqrt{\Delta\tau}x_{\gamma} \hat{v}_{\gamma}}
+
\mathcal{O}\left(\Delta\tau^{2}\right)\,.
\end{equation}
Combining Eqs.~\eqref{eq:primitive} and \eqref{eq:hs} yields the following representation of 
the short-time propagator,
\begin{equation}
\label{eq:hs2}
e^{-\Delta\tau\left(\hat{H}-\mathcal{E}_{0}\right)}
=
\int d{\bf x}\,p({\bf x})\,\hat{B}({\bf x})
+
\mathcal{O}\left(\Delta\tau^{2}\right)\,,\quad
\end{equation}
with $p({\bf x)} = (2\pi)^{-\frac{N_{\gamma}}{2}} \, e^{-\frac{|{\bf x}|^{2}}{2}}$ and
\begin{equation}
\hat{B}({\bf x})
=
\exp\left(-\Delta\tau\,\left(H_{0}-\mathcal{E}_{0}+\hat{v}_{0}\right)
+
\sqrt{\Delta\tau}\sum_{\gamma}x_{\gamma}\hat{v}_{\gamma}\right)\,.
\end{equation}
The short-time propagator is thus written as an integral over normally distributed parameters 
${\bf x}$, called auxiliary fields, of exponentials of one-body operators, $\hat B({\bf x})$. 
Eq.~\eqref{eq:hs2} maps the original interacting system onto an ensemble of non-interacting 
systems coupled with fluctuating external potentials. The integration over the auxiliary fields 
${\bf x}$ recovers the effect of the two-particle interaction.
While for a completely general two-body Hamiltonian the number of auxiliary fields is $N_\gamma =
\mathcal{O}(M^2)$
\cite{AlSaidi_JCP124_2006,Zhang_Notes_2013,Motta_JCP140_2014}, for real materials 
$N_\gamma$ can be brought down to $\mathcal{O}(M)$, an example of which is given next.

\subsubsection{Modified Cholesky decomposition}

To perform the HS transformation described above, we can rewrite the Hamiltonian as
\begin{equation}
\begin{split}
\label{eq:mcd1}
H
&=
H_{0}+\sum_{pq}\left(h_{pq}-\frac{1}{2}\sum_{r}v_{prrq}\right)
\, 
\sum_\sigma \crt{p\sigma} \dst{q \sigma} 
+ \\
&+ \frac{1}{2}\sum_{spqr}v_{pqrs} 
\, 
\sum_{\sigma\tau} \crt{p \sigma} \dst{r \sigma} \crt{q \tau} \dst{s \tau}\,.
\end{split}
\end{equation}
Representing (exactly or approximately) $v_{pqrs}$ as a sum $v_{pqrs} = \sum_\gamma 
L^{\gamma}_{pr} L^{\gamma}_{qs}$ of terms in which the indices $pr$, $qs$ are uncoupled 
immediately leads to the form \eqref{eq:hamsum}, with $\hat{v}_0$ defined in 
\eqref{eq:mcd1} and
\begin{equation}
\hat{v}_\gamma = i \sum_{pq} L^\gamma_{pq} \sum_{\sigma} \crt{p \sigma} \dst{q \sigma}\,.
\end{equation}
There are many possible ways of manipulating $v_{pqrs}$, and taking advantage of this 
flexibility is essential to improving the efficiency and accuracy of AFQMC.
When real-valued orbitals are used as one-electron basis set, the interaction tensor can 
be reshaped into a real-valued, symmetric and positive-definite matrix $V_{\mu(pr)\nu(qs)} 
= v_{pqrs}$, which can then be approximated using a modified Cholesky 
decomposition (mCD) algorithm \cite{Beebe_IJQC12_1977,Koch_JCP118_2003,Aquilante_JCC31_2010}.
The goal of the mCD is to  recursively reach a value $N_\gamma$ when 
the maximum error in representing $V_{\mu\nu}$ is less than 
a predetermined threshold value, $\delta$.
Given $N_\gamma \geq 0$ Cholesky vectors 
$\left\{ L_{\mu}^{\gamma}\right\} _{\gamma=1}^{N_\gamma}$, 
$V_{\mu\nu}$ can be expressed as
\begin{equation}
\label{eq:chol}
V_{\mu\nu}
=
\sum_{\gamma=1}^{N_\gamma}L_{\mu}^{\gamma}L_{\nu}^{\gamma}
+
\Delta_{\mu\nu}^{(N_\gamma)}\,,
\end{equation}
where $\Delta_{\mu\nu}^{(N_\gamma)}$ is the residual error matrix at the 
$N_\gamma$-th iteration. Let $\nu_{0}$ denote the index of the largest 
diagonal element of $\Delta_{\mu\nu}^{(N)}$:
\begin{equation}
\Delta_{\max} = \max_\nu \Delta_{\nu \nu}^{(N_\gamma)} 
= 
\Delta_{\nu_0\nu_0}^{(N_\gamma)}\,.
\end{equation}
A damped prescreening operation is performed, i.e., elements satisfying 
\begin{equation}
\sqrt{ |\Delta^{(N_\gamma)}_{\nu\nu} | \Delta_{\max} } \leq \min\{10^{-9},\delta\}
\end{equation}
are put equal to zero. 
As noted in Refs.~\cite{Koch_JCP118_2003,Aquilante_JCC31_2010}, the damping 
serves as a safeguard against rounding errors that may render the decomposition 
numerically unstable. The next-iteration Cholesky vector is then obtained as 
\begin{equation}
\label{eq:chol2}
L_{\mu}^{(N_\gamma+1)}
=
\frac{ \Delta_{\mu\nu_{0}}^{(N_\gamma)} }{ \sqrt{ \Delta_{\max} } }\,,
\end{equation}
and the residual error matrix is updated accordingly. Note that the positivity of $V$ 
and the recursion relation \eqref{eq:chol2} imply that $\Delta^{(N_\gamma)}$ is 
always non-negative.
When $\Delta_{\mu\nu}^{(N_{\gamma})} < \delta$, the Cauchy-Schwarz inequality implies 
$|\Delta_{\mu\nu}^{(N_{\gamma})} | \leq \sqrt{ \Delta_{\mu\mu}^{(N_{\gamma})} 
\Delta_{\nu\nu}^{(N_{\gamma})} } = \delta$ for all $(\mu,\nu)$, and the iteration terminates.

\begin{figure}[ht!]
\begin{center}
\includegraphics[width=0.45\textwidth]{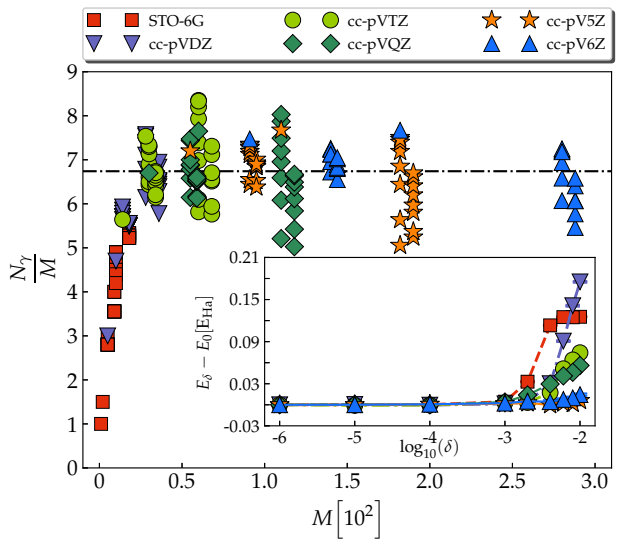}
\caption{(color online) 
The ratio $N_\gamma/M$ between Cholesky vectors and basis functions,
for atoms He to Kr and dimers He$_2$ to Kr$_2$ at bondlength $R = 2 \, a_B$, 
using STO-6G and cc-pVxZ bases with x=2,3,4,5,6 and threshold $\delta = 10^{-6}$. 
Asymptotically $N_\gamma \simeq 7 M$.
Inset: dependence of the AFQMC energy $E_\delta$ on the threshold 
$\delta$ for Cl$_2$ at experimental equilibrium bondlength, $R=3.7566$ 
$a_B$. $\delta \leq 10^{-5}$ yield energies within 1\,m$\mathrm{E_{Ha}}$ 
of converged value.
} 
\label{fig:cholesky}
\end{center}
\end{figure}

Typical values of $\delta$ range between $10^{-4}$ and $10^{-6}$. Performing a mCD 
decomposition of $V_{\mu\nu}$ requires $\mathcal{O}(M^{3})$ operations 
and memory \cite{Purwanto_JCP135_2011}, and leads to $\mathcal{O}(M)$ Cholesky 
vectors, as exemplified in Fig.~\ref{fig:cholesky}.
Direct diagonalization of $V_{\mu\nu}$, used in early implementations of AFQMC, leads 
instead to $\mathcal{O}(M^2)$ vectors, and requires $\mathcal{O}(M^6)$ operations and 
$\mathcal{O}(M^4)$ storage, which would be prohibitive for large molecules. 

\subsubsection{Mean-field subtraction}

Different forms of the HS transformation exist which can affect the performance of 
AFQMC. It is computationally advantageous to rewrite the Hamiltonian \eqref{eq:hamsum} 
prior to the HS decomposition, by subtracting an estimate of the ground-state average of 
each operator $\hat{v}_{\gamma}$, 
\begin{equation}
\begin{split}
\hat{H} 
&= 
H_{0}  
+ 
\sum_{\gamma} \frac{\overline{v}_{\gamma}^{2}}{2} 
+  
\hat{v}_{0}
- 
\sum_{\gamma}\overline{v}_{\gamma} \hat{v}_{\gamma}
-
\sum_{\gamma} \frac{\left(\hat{v}_{\gamma}-\overline{v}_{\gamma}\right)^{2}}{2} 
= \\
&= H_{0}^{\prime}
+
\hat{v}_{0}^{\prime}
-
\sum_{\gamma} \frac{\left(\hat{v}^\prime_{\gamma} \right)^{2}}{2}\,. 
\end{split}
\end{equation}
where the modified one-body operators are
 $\hat{v}^\prime_{\gamma}=\hat{v}_{\gamma}-\overline{v}_{\gamma}$, and the constant 
shift can be obtained, for example, using the trial wave function,
\begin{equation}
\overline{v}_{\gamma}
=
\frac{\langle\Psi_{T}|\hat{v}_{\gamma}|\Psi_{T}\rangle}{\langle\Psi_{T}|\Psi_{T}\rangle} \, .
\end{equation}
This is called mean-field background in the AFQMC literature. HS transformation applied 
to the modified one-body operators can lead to smaller fluctuations in free-projection 
calculations and smaller systematic errors when a constraint is applied 
\cite{AlSaidi_JCP124_2006,Purwanto_JCP130_2009}.

\subsection{The free-projection AFQMC}

With the formalism established so far, it is straightforward to realize the projection process 
by random walks in the space of $\mathcal{S}(N)$ and evaluate the ground-state energy: 
\begin{equation}
\begin{split}
\label{eq:freep}
\mathcal{E}(n\Delta\tau) 
&=
\frac
{\langle\Psi_{T}| \ham e^{-n\Delta\tau(\ham-\mathcal{E}_{T})}|\Psi_{I}\rangle}
{\langle\Psi_{T}|e^{-n\Delta\tau(\ham-\mathcal{E}_{T})}|\Psi_{I}\rangle}
= \\
&= \frac
{\int\prod_{l=0}^{n-1}d{\bf x}_{l}\,p({\bf x}_{l})\,W_{n}\,e^{i\theta_{n}}
\,
\frac{\langle\Psi_{T}|\ham|\Psi_{n}\rangle}{\langle\Psi_{T}|\Psi_{n}\rangle}
}{
\int\prod_{l=0}^{n-1}d{\bf x}_{l}\,p({\bf x}_{l})\,W_{n}\,e^{i\theta_{n}}
}\,,
\end{split}
\end{equation}
where $\Psi_{T}$ is a trial wavefunction, typically the SD from HF or DFT, or a linear 
combination of SDs from a complete-active space self-consistent field (CASSCF) 
calculation, for example. For convenience, $\Psi_{T}$ is often chosen to be the same 
as $\Psi_{I}$, although this is not required. (We can take advantage of the flexibility 
to choose them differently in order to help impose certain symmetry properties in the 
calculations, see below and Refs. \cite{Purwanto_JCP128_2008,Shi_PRB88_2013}).
We have written 
$|\Psi_{n}\rangle=\hat{B}({\bf x}_{n-1})... \hat{B}({\bf x}_{0})|\Psi_{I}\rangle$ 
and
$W_{n}\,e^{i\theta_{n}}=\langle\Psi_{T}|\Psi_{n}\rangle$ to reduce clutter.

One way to evaluate the integrals in Eq.~\eqref{eq:freep} is by sampling a fixed-length 
(of $n$-segments) path in the auxiliary-field space, for example, by the Metropolis Monte 
Carlo (MC) algorithm. This is the form in which the auxiliary-field approach was first proposed, 
both at finite-temperature \cite{Blankenbecler_PRD_1981} 
and at $T=0\,K$ \cite{Sugiyama_Annals_1986}, in model systems.
Here we will describe a different  approach by casting \eqref{eq:freep} into branching, 
open-ended random walks in the space of SDs.
For free-projection calculations per se, there is no particular advantage in using this 
approach over the more standard Metropolis procedure; in fact, it is more convenient to 
compute observables or correlations with the latter. 
However, in order to impose a constraint to control the sign problem (and phase problem) 
as we will discuss next, there is a fundamental difference in how the two
algorithms scale, and it was necessary to recast the process as open-ended random 
walks \cite{Zhang_PRL_1995}. 

In the random walk realization of Eq.~\eqref{eq:freep}, the wavefunction 
$e^{-n\Delta\tau(\ham-\mathcal{E}_{T})}|\Psi_{I}\rangle$ is represented, in a MC sense, 
by an ensemble of SDs, $|\Psi_{n,w}\rangle$ with weights $W_{n,w}$ and phases 
$e^{i\theta_{n,w}}$.
The structure $\{ \Psi_{n,w},W_{n,w},\theta_{n,w}\}$ is called a random walker.
The calculation begins by initializing a population of $N_{w}$ walkers, 
at $\left\{ \Psi_{I},1,0 \right\}$.
The population size $N_w$ is either held fixed at 
a preset 
value or allowed to fluctuate around it throughout the calculation, applying population control as needed. 

As the random walk proceeds, for all walkers $w$ and times $k$, we sample an auxiliary 
field ${\bf x}_{k,w}$ and update the walker state, weight and phase as 
\begin{equation}
\begin{split}
\label{eq:weight_update}
|\Psi_{k+1,w} \rangle 
&= \hat{B} ({\bf x}_{k,w}) |\Psi_{k,w} \rangle \,,  \\
W_{k+1,w} \, e^{i\theta_{k+1,w}} 
&=
\frac{ \langle \Psi_{T} | \Psi_{k+1,w} \rangle }{ \langle \Psi_{T} | \Psi_{k,w} \rangle }\,W_{k,w} \, e^{i\theta_{k,w} }\,.
\end{split}
\end{equation}
By Thouless' theorem, $|\Psi_{k+1}\rangle$  is a SD if $|\Psi_{k}\rangle$ is a SD,
so that all walkers maintain simple structure and efficient parametrization during 
the random walk. The ground-state energy can be computed as weighted average 
of the local energy functional,
\begin{equation}
\label{eq:localenergy}
\mathcal{E}_{0}
\simeq
\frac{ \sum_{w} W_{k,w} \, e^{i\theta_{k,w} } \mathcal{E}_{loc}(\Psi_{k,w}) }
{ \sum_{w} W_{k,w} \, e^{i \theta_{k,w} } }
\,\, , \,\,
\mathcal{E}_{loc}(\Psi)=\frac{\langle\Psi_{T}|\ham| \Psi\rangle}{\langle\Psi_{T}| \Psi\rangle}\,.
\end{equation}

\begin{figure*}[ht!]
\includegraphics[width=0.45\textwidth]{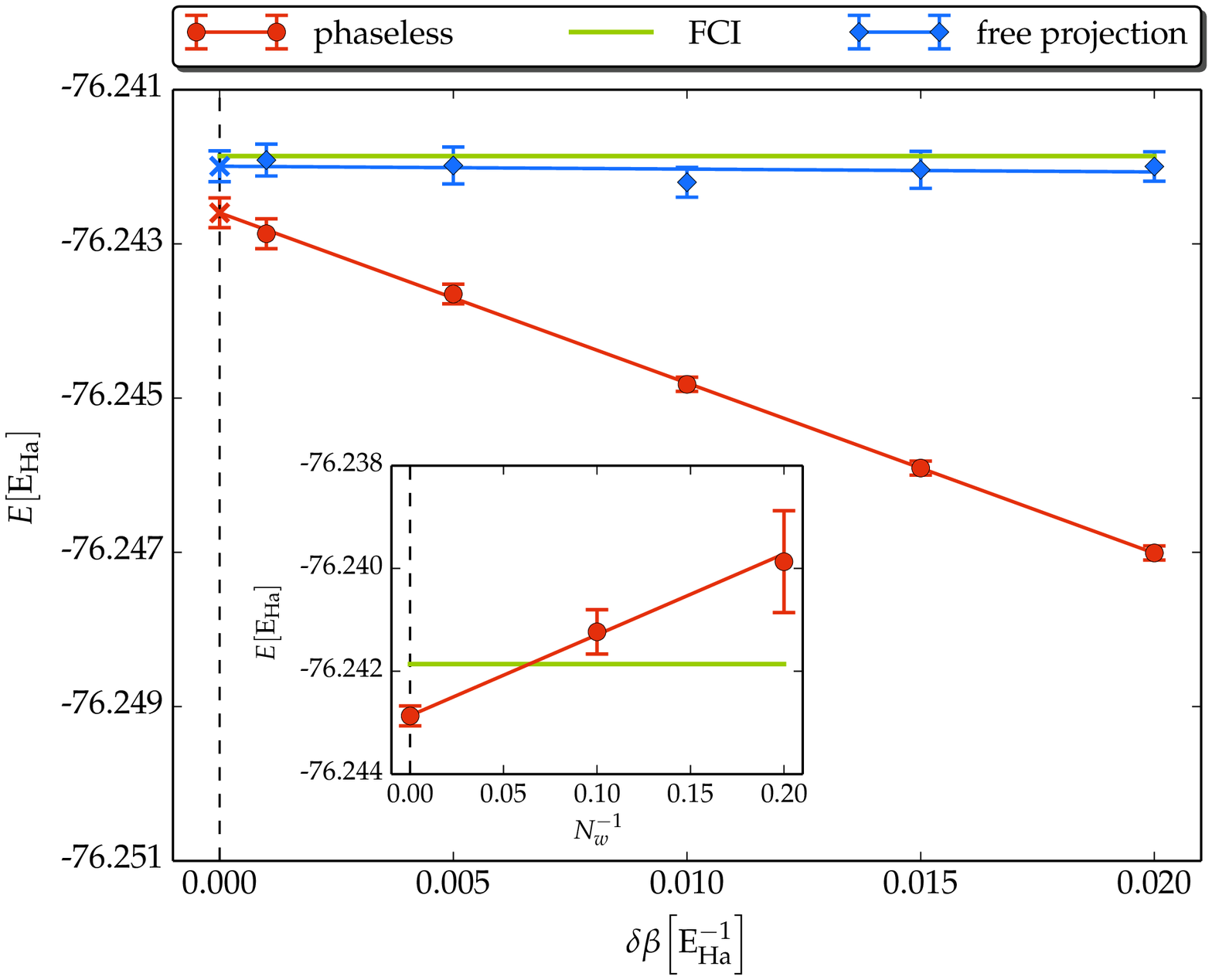}
\includegraphics[width=0.45\textwidth]{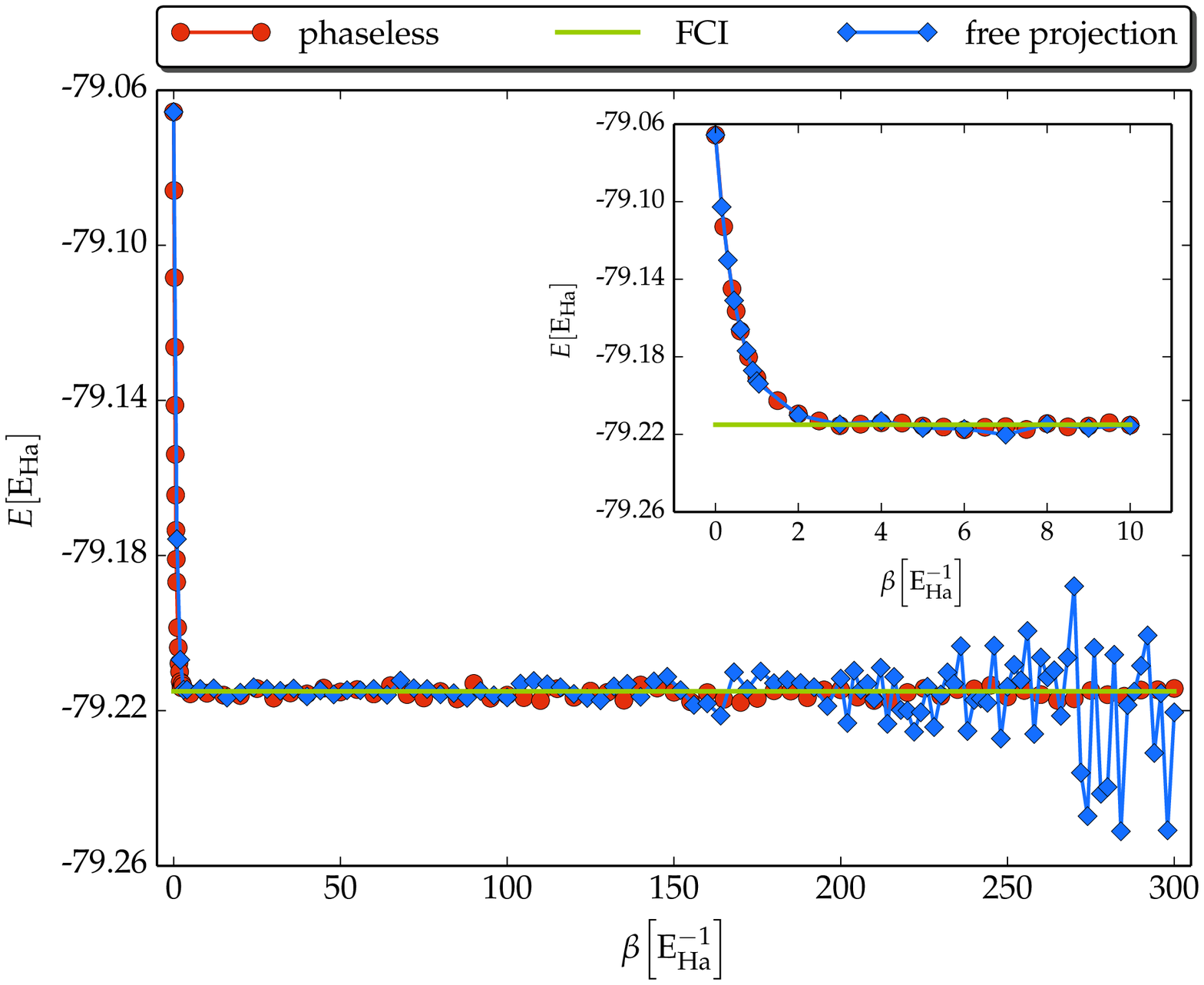}
\caption{(color online) Left: time step extrapolation of free-projection and phaseless AFQMC 
energies for H$_2$O (cc-pVDZ level, triangular geometry with $R_{OH} = 1.8434 \, \mathrm{a_B}$, 
$\theta_{HOH} = 110.6^o$), using a population of $N_w = 2 \times 10^4$ walkers. 
(inset: extrapolation of phaseless AFQMC energies vs.~the inverse population size, illustrating 
population control bias at small $N_w$.
A time step of $\Delta\tau = 10^{-3} \mathrm{E_{Ha}^{-1}}$ was used here.)
Right: emergence of the phase problem in free-projection AFQMC calculations of ethane (STO-6G 
basis). (inset: comparison between free-projection and phaseless AFQMC for short projection time).
} 
\label{fig:extrapolation}
\end{figure*}

The free-projection AFQMC estimator of the ground-state energy is exact. The walkers 
will need to be periodically re-orthonormalized to maintain numerical precision. The only 
potential systematic errors come from the Trotter error of finite time-step, and population 
control bias. These can be easily removed by standard extrapolation procedures, 
as illustrated in Fig.~\ref{fig:extrapolation}.
The Trotter error in free-projection is typically quadratic in $\Delta\tau$. 
Population control bias is typically $1/N_w$, and should be small in a free-projection calculation, since the population 
size is necessarily large; however, the calculation is prone to large fluctuations in the 
weights because of intrinsic instability from the sign/phase problem.

A detailed flowchart of the algorithm will be presented after constrained AFQMC calculations
are introduced. Additional implementation details, 
including population control and re-orthonormalization of the walkers, which are shared
by free-projection and constrained calculations,
 will be discussed under Implementation issues.

\section{Constrained AFQMC calculations} 

\subsection{The phase problem}

The free-projection AFQMC suffers from an asymptotic instability in $n\Delta\tau$, 
which is the manifestation of the fermion sign problem 
\cite{Feynman_book_1965,Loh_NATO41_1990} in many model systems such as the 
doped repulsive Hubbard model. For more general interactions, such as the long-range Coulomb  
repulsion, a more severe instability appears  \cite{Zhang_PRL90_2003}, which has been referred 
to as the phase problem in AFQMC literature.
More specifically, in free-projection AFQMC, $\mathcal{E}(n\Delta\tau)$ is estimated by 
a ratio whose denominator is the expectation $\langle \vartheta \rangle_\Psi$, with 
$\vartheta\equiv e^{i\theta(\Psi)}$, with respect to the random variable $\Psi$ 
over the manifold of SDs. (The numerator is given by $\langle \vartheta\,\mathcal{E}_{loc}(\Psi)
 \rangle_\Psi$.)
Unless prevented by special symmetry, the expectation of the phase approaches zero 
exponentially with $n\Delta\tau$. This causes the variance in the MC estimator of the 
energy to grow exponentially, $\sigma_{n}\simeq e^{\gamma n\Delta\tau}$.  The rate 
$\gamma$ at which the variance increases depends on the nature of the interaction.
For each system, the rate grows with $M$ and $N$, 
making the instability of free-projection AFQMC more severe for large basis sets and 
systems. As the Central Limit Theorem dictates, computing the average $\mathcal{E}
(n\Delta\tau)$ to a desired accuracy $\epsilon^{*}$ requires the number  
of samples to scale as $N_{w}\simeq\frac{\sigma_{n}^{2}}{\epsilon^{*}} = e^{2\gamma 
n\Delta\tau}/\sigma^{*}$. 
The exponential increase of noise with imaginary time translates into a formal exponential 
scaling of free-projection AFQMC. The exponential growth in MC noise from the phase 
problem is illustrated in Fig.~\ref{fig:extrapolation}.

To gain a sense of how the phase problem arises, let us recall that we are seeking a 
representation of the solution in Eq.~\eqref{eq:togs} in the form
\begin{equation}
\label{eq:stochastic1}
|\Psi(\beta)\rangle = \int_{ \mathcal{S}(N) } d\Psi \, f_\beta(\Psi) \, |\Psi\rangle\,,
\end{equation}
in the over-complete space of non-orthogonal SDs, where the coefficients $f_\beta(\Psi)$ 
are represented stochastically by the probability distribution of MC samples of $\Psi$ 
\cite{Fahy_PRL65_1990,Fahy_PRB43_1991}. 
The manifold of SDs contains both a generic determinant $\Psi$ and its opposite, 
$- \Psi$, as distinct points. Indeed it contains $e^{i\theta}\,\Psi$ for any $\theta\in[0,2\pi)$. 
This symmetry property has its root in the linearity of the Schr\"odinger 
equation: if $|\Psi(\beta)\rangle$ is a solution to Eq.~\eqref{eq:imt}, so is $-|\Psi(\beta)\rangle$; 
similarly, if $f_\beta(\Psi)$ is a solution in Eq.~\eqref{eq:stochastic1} defined on a 
domain in the SD manifold comprising $\{ \Psi\}$, so is $f_\beta(e^{i\theta}\,\Psi)=f_\beta(\Psi)$, 
defined on a domain comprising $\{ e^{i\theta}\,\Psi\}$. 

In the framework of a numerical simulation, the symmetry between equivalent domains in 
$\mathcal{S}(N)$ 
implies that, for any statistical ensemble of walkers 
giving an MC representation of the ground-state wavefunction, there exist other ensembles 
providing equivalent representations. If $\hat{B} ({\bf x})$ is real, there are two equivalent 
ensembles, related by a sign change. If  $\hat{B} ({\bf x})$ is complex, there are infinitely 
many ensembles given by different gauge choices.

Apart from special cases, such as the half-filled repulsive Hubbard model 
\cite{Hirsch_PRB31_1985}, where particle-hole symmetry yields a $\hat{B} ({\bf x})$ that 
confines the random walk in one domain with a single choice of sign (either $+$ or $-$), 
the random walks will in general reach other domains. 
The MC dynamics given by $\hat{B} ({\bf x})$ does not have any means to detect the crossing 
of ``boundaries'' between different domains: the walkers evolve continuously in the random 
walk \cite{Zhang_Notes_2013}; because of the high dimensionality of $S(N)$, 
we cannot tabulate a history of where the random walk has 
visited to estimate a sign or phase 
(in the spirit of cancellation algorithms \cite{Zhang_PRL67_1991,Anderson_JCP95_1991,Booth_JCP_2009}),
without incurring exponential scaling  \cite{Zhang_Notes_2013}.
The exploration of $\mathcal{S}(N)$ being ergodic, in the $\beta\to\infty$ limit the MC representation 
of the ground-state consists of an equal mixture of walkers from degenerate domains, regardless 
of whether the initial distribution was concentrated in one or not.
This  symmetry results in an exponentially fast decay in the signal-to-noise ratio, i.e. in the
average sign or phase, given by the expectation value $\langle \vartheta\rangle_\Psi$.

In this framework,  control of the sign problem can be achieved by modifying the dynamics 
so as to break the symmetry in the imaginary-time evolution or, equivalently, to prevent Slater 
determinants from crossing into other domains. 

For real $\hat{B} ({\bf x})$, the necessary symmetry breaking is achieved by deleting walkers 
whose overlap with the trial wavefunction $|\Psi_T\rangle$ changes sign, i.e., when $\Psi$ 
crosses the hyper-surface  $\omega(\Psi)\equiv\langle \Psi_T|\Psi\rangle=0$. The resulting
approximation defines the constrained-path AFQMC (cp-AFQMC) method, which has delivered 
excellent results for lattice models of correlated electrons \cite{LeBlanc_PRX5_2015}.

The phase problem affecting AFQMC calculations for real materials emerges as the 
Hubbard-Stratonovich transformations result in complex-valued $\hat{v}_{\gamma}$ 
operators. (It is possible to maintain real values but these were found to have much 
larger fluctuations \cite{Zhang_PRL90_2003}).
The constraints used to remove that instability required an additional step beyond the 
constrained-path approximation.
The approach to real materials has been referred to in the literature as the phaseless 
or phase-free approximation. Once the latter is formulated, the former can be obtained 
as a special case.
Here in discussing both these situations, where a constraint is applied on the sampling 
of the auxiliary-field paths in the random walk, we will often refer to them as cp-AFQMC 
for simplicity. Below we will introduce the key ingredients that enable the formulation of 
the cp-AFQMC algorithm.

\subsection{Importance sampling transformation}

Importance sampling is a variance reduction technique widely used in classical and 
quantum Monte Carlo \cite{Hammersley_RS16_1954,Rosenbluth_JCP23_1955}.
Within AFQMC, importance sampling was introduced in the case of real propagators 
$\hat{B} ({\bf x})$ as a variance-reduction measure \cite{Zhang_PRB55_1997,Zhang_PRL_1999}.
For complex $\hat{B} ({\bf x})$,  the importance sampling transformation is applied not just
in the usual sense of variance reduction, but to define a gauge condition \cite{Zhang_PRL90_2003}. 

We will outline the importance-sampling transformation for the most general case below. 
The case of real-valued $\hat{B} ({\bf x})$, such as in the Hubbbard model, will follow as 
a special example.
Recalling that the Hubbard-Stratonovich transformation is defined up to a complex-valued 
shift $\overline{{\bf x}}$\cite{Zhang_PRL90_2003,Purwanto_PRE70_2004}, we can rewrite
\begin{equation}
\mathcal{E}(n\Delta\tau)
\simeq
\frac
{\int\prod_{l=0}^{n-1}d{\bf x}_{l}\,p({\bf x}_{l}-\overline{{\bf x}}_{l})\,\langle\Psi_{T}| \ham |\Psi_{n}\rangle}
{\int\prod_{l=0}^{n-1}d{\bf x}_{l}\,p({\bf x}_{l}-\overline{{\bf x}}_{l})\,\langle\Psi_{T}|\Psi_{n}\rangle} \, ,
\end{equation}
where now $|\Psi_{n}\rangle = \hat{B}({\bf{x}}_{n-1}-\overline{\bf{x}}_{n-1}) \dots 
\hat{B}({\bf{x}}_{0}-\overline{\bf{x}}_{0}) |\Psi_I \rangle$. Formally, we can apply a similarity 
transformation in each step along the random walk path using the overlap ratio 
$\langle \Psi_T|\Psi_{l+1}\rangle/\langle \Psi_T|\Psi_{l}\rangle$, to obtain an equivalent form:
\begin{equation}
\mathcal{E}(n\Delta\tau)
\simeq
\frac
{\int\prod_{l=0}^{n-1}d{\bf x}_{l}\,p({\bf x}_{l})\,I({\bf x}_{l},\overline{{\bf x}}_{l};\Psi_{l})
\,
\frac{\langle\Psi_{T}| \ham |\Psi_{n}\rangle}{\langle\Psi_{T}|\Psi_{n}\rangle}
}{
\int\prod_{l=0}^{n-1}d{\bf x}_{l}\,p({\bf x}_{l})\,I({\bf x}_{l},\overline{{\bf x}}_{l};\Psi_{l})
} \,,
\end{equation}
where the importance function is defined as
\begin{equation}
\label{eq:hybrid-wt}
I({\bf x},\overline{{\bf x}};\Psi)
=
\frac
{\langle\Psi_{T}| \hat{B}({\bf x}-\overline{{\bf x}})|\Psi\rangle}
{\langle\Psi_{T}|\Psi\rangle}
\,
\exp
\left({\bf x}\cdot\overline{{\bf x}}-\frac{\overline{{\bf x}}\cdot\overline{{\bf x}}}{2}\right)\,. 
\end{equation}
Note that the overlap ratio applied in our similarity transformation is in general 
complex, which is different from  standard importance-sampling transformations. 
Introducing a short-hand for the local ``mixed'' estimator:
$\frac{\langle\Psi_{T}| \hat{A} |\Psi\rangle}{\langle\Psi_{T}|\Psi\rangle}\equiv
\langle \hat{A} \rangle_m$, and restricting our (arbitrary) choices of 
$\overline{{\bf x}}$ to be 
analytic functions of $\Delta\tau^{\frac{1}{2}}$, we arrive at the following 
\begin{equation}
\log I({\bf x},\overline{{\bf x}};\Psi)
=
\sum_{\gamma}x_{\gamma}\left(\overline{x}_{\gamma}
+
\sqrt{\Delta\tau}\,\langle \hat{v}_{\gamma}\rangle_m\right)
+
\mathcal{O}(\Delta\tau)
\end{equation}
by expanding $\log I({\bf x},\overline{{\bf x}};\Psi)$ in a Taylor series. 
It is evident that the choice 
\begin{equation}
\label{eq:fb}
\overline{x}_{\gamma}=-\sqrt{\Delta\tau}\langle \hat{v}_{\gamma}\rangle_m
\end{equation}
cancels fluctuations in the importance function to first order in $\Delta\tau^{\frac{1}{2}}$, leading 
to a more stable random walk. The dynamic shift, which we will refer to as a \emph{force bias},
modifies the MC sampling of the auxiliary-fields by shifting the center of the Gaussian (or Bernoullian
in the case of discrete Ising fields \cite{Shi_PRB88_2013})  according to the overlap function $\langle \Psi_T| \Psi\rangle$. 
For complex $ \hat{v}_\gamma$, in general the force bias  has an imaginary component. 
This is connected to the complex character of the overlap function in the similarity transformation discussed 
above, and turns out to play a key role for obtaining an optimal  constraint to control the phase problem. 

Inserting the optimal force bias into $I({\bf x},\overline{{\bf x}};\Psi)$ and expanding 
$\log I({\bf x},\overline{{\bf x}};\Psi)$ to first order in $\Delta\tau$ yields the expression
\begin{equation}
\begin{split}
\label{eq:I-El-intermediate}
&\log I({\bf x},\overline{{\bf x}};\Psi)=-\Delta\tau\left(\langle \ham \rangle-\mathcal{E}_{0}\right) + \\
&\frac{\Delta\tau}{2}
\sum_{\gamma}
\Big( 
\delta_{\gamma\gamma^{\prime}}-x_{\gamma}x_{\gamma^{\prime}}
\Big)\Big(
\langle \hat{v}_{\gamma}\hat{v}_{\gamma^{\prime}}\rangle
-
\langle \hat{v}_{\gamma}\rangle\langle \hat{v}_{\gamma^{\prime}}\rangle
\Big)
+
\mathcal{O}\left(\Delta\tau^{\frac{3}{2}}\right)\,.
\end{split}
\end{equation}
Since $\int d{\bf x}\,p({\bf x})\,x_{\gamma}x_{\gamma^{\prime}}=\delta_{\gamma\gamma^{\prime}}$ 
the last term of \eqref{eq:I-El-intermediate} has zero average over auxiliary fields, leading to 
the approximate expression 
\begin{equation}
\label{eq:importance_sampling}
I({\bf x},\overline{{\bf x}};\Psi)\simeq e^{-\Delta\tau\left(\mathcal{E}_{loc}(\Psi)-\mathcal{E}_0 \right)}
\end{equation}
for the importance function. 
This  is often referred to as the local energy formalism while Eq.~\eqref{eq:hybrid-wt} is referred 
to as the hybrid formalism.

By the introduction of the force bias, the importance sampling transformation guides the sampling 
of walkers away from regions of $\mathcal{S}(N)$ where $\omega(\Psi) \simeq 0 $, as  $\overline{{\bf x}}$ 
diverges. In the case when  $\hat{v}_{\gamma}$ is real, the procedure, taken at the limit of $\Delta
\tau\rightarrow 0$, imposes an infinite barrier at $\omega(\Psi) \simeq 0 $, and prevents the random 
walk from entering the other domain, thereby removing the sign problem.
In the case of a phase problem, however, the transformation and the force bias alone do not lead 
to complete control of the phase problem. Achieving this goal requires combining importance sampling 
with the phaseless approximation described next.
  
\subsection{The phaseless approximation}

From the expressions of the local energy and importance functions,
\eqref{eq:localenergy} and \eqref{eq:importance_sampling}, it is easy to see the role of $\omega(\Psi)=0$
in the phase problem:  both quantities involve fractions with denominator $\omega(\Psi)$, which diverges 
at the origin of the complex plane causing large fluctuations in AFQMC estimators.
As mentioned, in the presence of complex-valued $\hat{v}_{\gamma}$ operators, the random walk dynamics is 
invariant under the transformations $\Psi \to e^{i\theta} \Psi$, $\theta \in [0,2\pi)$.
As the projection time grows, the overlaps $\omega(\Psi)$ undergo a rotationally-invariant random walk 
in the complex plane, resulting in a finite concentration of walkers at the origin \cite{Zhang_Notes_2013}.

The phase problem affecting AFQMC for real materials is unique, in that not only does the weight 
of the walker acquire a phase, but the orbitals of the SD become complex. 
Straightforward generalizations of the constrained path approximation, e.g., imposing the condition  
$\mbox{Re}\,\omega(\Psi)>0$ without the use of a complex force bias, lead typically to poor results
\cite{Zhang_PRL90_2003}.
The complex force bias of Eq.~\eqref{eq:fb} keeps the overall phase stable to leading order 
in $\Delta\tau$, thereby allowing a particular overall gauge to be chosen to break the rotational 
symmetry.
 
In the framework of electronic structure AFQMC calculations, control of the phase problem is 
achieved by the real local energy and the phaseless approximations. The real local 
energy approximation replaces the $\mathcal{E}_{loc}(\Psi)$ term in the importance function 
with its real part, thereby leading to real and positive weights. 
The impact of this approximation was found to be typically mild, and the imaginary part of the local 
energy can often be carried for extended projection time, with little effect on the computed 
energy. (See below on the effect for back-propagation and observables \cite{Motta_backpropagation_2017},
however.)
The phaseless approximation aims to break the rotationally invariance of the random walk 
undergone by the overlaps $\omega(\Psi)$ by projecting them onto an adaptively evolving straight 
line. The projection is accomplished by changing the importance function as follows:
\begin{equation}
\label{eq:if_rle}
I({\bf x},\overline{{\bf x}};\Psi)
\simeq 
e^{-\Delta\tau\left(\mbox{Re}\mathcal{E}_{loc}(\Psi)-\mathcal{E}_{0}\right)}
\times
\max\left(0,\cos\left(\Delta\theta\right)\right)\,, 
\end{equation}
where $\Delta\theta=\mbox{Arg} \, \frac{\omega(\Psi^{\prime})}{\omega(\Psi ^{\phantom{\prime}})}$
is the phase difference between the overlaps, with the new walker state given by
$| \Psi ^{\prime}\rangle=\hat{B}({\bf x}-\overline{{\bf x}})| \Psi\rangle$. 
The factor $\cos\left(\Delta\theta\right)$ prevents stochastic trajectories from undergoing abrupt 
phase changes, and ensures that the probability distribution of $\omega(\Psi)$ vanishes at the origin.

With the optimal complex force bias, the '$\cos$' form can be replaced by other choices which yield similar results 
\cite{Zhang_CPC169_2005}. Notice that, when the $\hat{v}_\gamma$ operators are real-valued, 
the projection reduces to the constrained-path AFQMC, because $\Delta\theta =  0$ or $\pi$ and 
$\max\left(0,\cos\left(\Delta\theta\right)\right) = 1,0$, respectively. 

The resulting method, the phaseless AFQMC (ph-AFQMC), is the state-of-the-art technique for 
performing AFQMC calculations in real materials. A ph-AFQMC calculations differs from a 
fp-AFQMC in the update of walkers and weights, which is carried out as follows:
\begin{equation}
\begin{split}
\label{eq:weight_update_2}
|\Psi_{k+1,w} \rangle
&= \hat{B} ({\bf x}_{k,w}-\overline{{\bf x}}_{k,w}) |\Psi_{k,w} \rangle \, , \\
W_{k+1,w}
&= I({\bf x}_{k,w},\overline{{\bf x}}_{k,w};\Psi_{k+1,w})\, W_{k,w}  
\end{split}
\end{equation}
with the force bias and importance functions given by \eqref{eq:fb}, \eqref{eq:if_rle}, and 
$\theta_{k,w} \equiv 0$.

\subsection{Implementation issues}

\subsubsection{The trial wavefunction}

The choice of $\Psi_T$ is important in AFQMC. First, the closer $\Psi_T$ is to the ground state, 
the smaller the fluctuations in the local energy (if $\Psi_T = \Phi_0$, $\mathcal{E}_{loc}(\Psi) 
\equiv \mathcal{E}_0$).
In constrained calculations, $\Psi_T$ also determines the efficiency of importance sampling 
and the accuracy of the phaseless constraint.
Often using a single Slater determinant, such as the HF or DFT wavefunction,
delivers results of excellent accuracy. Use of multi-determinant trial wavefunctions,
\begin{equation}
| \Psi_T \rangle = \sum_{d = 1}^{N_d} A_d | \Psi_d \rangle\,, 
\end{equation}
can improve accuracy, reduce local energy fluctuations and the imaginary time needed to reach 
equilibration of walker ensembles.

Theorem \ref{thm:slater} can be straightforwardly generalized to multideterminant trial-wavefunctions, 
by treating each configurations $\Psi_d$ separately in the linear combination. 
Overlaps and Green's functions can be computed
as $\langle \Psi_T | \Psi \rangle = \sum_{d = 1}^{N_d} A^*_d \langle \Psi_d | \Psi \rangle$ and
\begin{equation}
\label{eq:usesherman}
G^{\Psi_T \Psi}_{p\sigma,q\tau} =\frac{ \sum_{d = 1}^{N_d} A^*_d \, \langle \Psi_d | \Psi \rangle \, 
G^{\Psi_\alpha \Psi}_{p\sigma,q\tau} }{ \sum_{d = 1}^{N_d} A^*_d \, \langle \Psi_d | \Psi \rangle }\,. 
\end{equation}
For many practical purposes, it is useful to write 
\begin{equation}
\label{eq:precomputing}
G^{\Psi_\alpha \Psi}_{p\sigma,q\tau} = \delta_{\sigma\tau} \, 
\sum_{r=1}^{N_\sigma} \Theta^{d}_{\sigma,qr} \, \left(V_{d\sigma}\right)^\dagger_{rp}\,, 
\end{equation}
where $V_{d\uparrow}$ and $V_{d \downarrow}$ are the matrices parametrizing $\Psi_d$.
For 
multideterminants from many standard QC approaches, such as CASSCF, the sum over $d$ can be greatly 
accelerated using the Sherman-Morrison-Woodbury formula \cite{Sherman_AMS20_1949,Press_Book_2007}.

\subsubsection{Propagation of walkers}

Updating a walker requires computing the force bias $\overline{\bf{x}}$. 
The cost of the operation can be brought down to $\mathcal{O}(N_\gamma N_d NM)$ 
by taking advantage of \eqref{eq:precomputing},
\begin{equation}
\label{eq:precomputing2}
\overline{x}_\gamma = - i \sqrt{\Delta\tau} \, 
\frac{ \sum_{d = 1}^{N_d} A^*_d \, \langle \Psi_d | \Psi \rangle \, 
\sum_\sigma \mbox{Tr} \left[ \mathcal{L}^\gamma_{d\sigma} \Theta^{d}_{\sigma} \right] 
}
{ \sum_{d = 1}^{N_d} A^*_d \, \langle \Psi_d | \Psi \rangle }\,, 
\end{equation}
where the tensor
$\left( \mathcal{L}^\gamma_{d\sigma} \right)_{rq} = \, 
\sum_p \left( V_{d\sigma} \right)^\dagger_{rp} L^\gamma_{pq}$ can be precomputed at 
the beginning of the simulation, and the trace in \eqref{eq:precomputing2}  costs 
$\mathcal{O}(NM)$ operations. (We apply a cap to  the force bias values if necessary
\cite{Purwanto_PRB80_2009} as discussed in the Appendix.)

Once the force bias is determined, the matrix 
$A({\bf{x}})_{pq} = - \Delta\tau v_{0,pq} + 
i \sqrt{\Delta\tau} \, \sum_\gamma (x_\gamma-\overline{x}_\gamma) L^\gamma_{pq}$ 
is computed, at the cost of $\mathcal{O}(N_\gamma M^2)$ operations, 
and its exponential is applied to the matrices $U_\sigma$ parametrizing $\Psi$.
Computing $e^{A({\bf{x}}) } U_\sigma$ as truncated Taylor series requires $\mathcal{O}
(M^2 N)$ operations.

\subsubsection{Local energy evaluation}

The local energy 
is needed for computing the ground-state energy, and it also controls the weights in constrained 
calculations. The most demanding part of its calculation comes from the interaction 
term, which via  Wick's theorem can be written as
\begin{equation}
\begin{split}
\label{eq:localenergy2}
&\frac{ \langle \Psi_T | \hat{H}_2 | \Psi \rangle }{ \langle \Psi_T | \Psi \rangle }
=
\sum_d
\frac{A^*_d \, \langle \Psi_d | \Psi \rangle}{ \sum_d A^*_d \, \langle \Psi_d | \Psi \rangle } \times \\
&\times \sum_\gamma \left[\sum_{pqrs}
L^\gamma_{pr} L^\gamma_{qs} 
\sum_{\sigma\tau} \left( 
G^{\Psi_\alpha \Psi}_{p\sigma,r\sigma} G^{\Psi_\alpha \Psi}_{q\tau,s\tau}
-
G^{\Psi_\alpha \Psi}_{p\sigma,s\tau} G^{\Psi_\alpha \Psi}_{q\tau,r\sigma} 
\right)\right]\,. 
\end{split}
\end{equation}
The scaling of \eqref{eq:localenergy2} can be brought down to $\mathcal{O}(N_d 
N_\gamma N^2 M)$. Indeed, a simple calculation shows that the expression inside 
the square bracket in Eq.~\eqref{eq:localenergy2}, the contribution to the local energy 
from the Cholesky vector $\gamma$ and the configuration $d$ in $|\Psi_T\rangle$ 
can be written as
\begin{equation}
\label{eq:localenergy3}
\Big( \sum_{\sigma} \mbox{Tr}\left[ \mathcal{L}^\gamma_{d\sigma} \Theta^{d}_\sigma \right] \Big)^2
- \sum_\sigma \mbox{Tr}\left[ \mathcal{L}^\gamma_{d\sigma} \Theta^{d}_\sigma 
\mathcal{L}^\gamma_{d\sigma} \Theta^{d}_\sigma \right]\,.
\end{equation}
The cost of the calculation results from computing on-the-fly the tensors
$\mathcal{L}^\gamma_{d\sigma} \Theta^{d}_\sigma$, requiring $\mathcal{O}(N_d N_\gamma N^2 M)$ 
operations, and the traces in \eqref{eq:localenergy3}, requiring $\mathcal{O}(N_d N_\gamma (N+N^2))$ 
operations. Additional reductions can be achieved for CASSCF type of $|\Psi_T\rangle$. 
(A cap can be placed on the local energy to regulate spurious fluctuations 
\cite{Purwanto_PRB80_2009} as discussed in the Appendix.)

\subsubsection{Population control}

As the random walk proceeds, 
some walkers accumulate very large weights while some obtain very small weights. These different 
weights cause a loss of sampling efficiency because the algorithm spends a disproportionate amount 
of time keeping track of walkers that contribute little to the energy estimate. 
To eliminate such inefficiency, a branching scheme is introduced to redistribute weights without 
changing their statistical distribution. In such a scheme, walkers with large weights are replicated 
and walkers with small weights are eliminated with appropriate probability.
However, because branching might cause the total population to fluctuate in an unbounded way 
(e.g. to grow to infinity or to perish altogether), we perform population control periodically to eliminate 
this instability. The population control can incur a bias when the total weight of the walkers is modified, 
as illustrated in Fig.~\ref{fig:extrapolation}. There exist approaches to reduce this bias by carrying 
a history of overall weight correction factors \cite{Calandra_PRB58_1998,Zhang_PRB55_1997,Umrigar_JCP99_1993}.
In typical cp-AFQMC calculations, population sizes of order hundreds or thousands are used,
and the population control bias is often negligible. 

\subsubsection{Numerical stabilization}

Repeated multiplications of the matrices $e^{A({\bf x})}$ onto the walker SD leads to accumulation 
of round-off errors and a loss of numerical precision. This introduces
spurious components in the walkers' orbitals which eventually overwhelm the 
exact components. The instability may be interpreted as a tendency of walkers to collapse to a 
Bose ground state, in which all particles occupy the lowest-energy orbital.
Within AFQMC, the instability is eliminated by periodically orthogonalizing the orbitals with, e.g.,
a modified Gram-Schmidt (mGS) procedure, by sweeping through the orbitals $\{u_i\}_{i=1}^{
N_\uparrow}$ and $\{v_i\}_{i=1}^{N_\downarrow}$ of $\Psi$
with the following steps:
\begin{equation}
|\tilde u^\prime_k \rangle = 
 |u_k \rangle - \sum_{j=1}^{k-1}\langle u^\prime_j | u_k \rangle\,
|u^\prime_j\rangle; \quad
|u^\prime_k \rangle = \frac{|\tilde u^\prime_k \rangle}{\sqrt{\langle  \tilde u^\prime_k |  \tilde u^\prime_k \rangle}}
\quad  ,
\end{equation}
for $k=1\dots N_\uparrow $,
and similarly for the spin-down orbitals. The mGS does not affect force biases or estimators 
of physical properties, 
thanks to Theorem 1.5.
We comment that, in coordinate space methods such as diffusion MC or Green's function Monte 
Carlo (GFMC), a similar tendency is present which is in fact the primary source of the sign 
problem in that framework. The procedure to stabilize the SDis in AFQMC is analogous to the
cancellation of walkers in GFMC \cite{Zhang_PRL67_1991,Anderson_JCP95_1991}, 
which requires a large density of walkers and does not scale well with system size. The structure 
of the SD allows the cancellation of the one-particle instability against the Bose state to be carried 
out analytically in AFQMC.

\subsection{The AFQMC algorithm}

\begin{enumerate}
\item Specify the initial state of each walker $\Psi$, and assign it a weight $1$ and phase $0$.
\item If the walker weight is non-zero, compute its overlap with the trial wavefunction and force bias.
\item Sample an auxiliary-field configuration ${\bf{x}}$ and propagate the walker in imaginary time.
\item Compute the overlap between the updated walker $\Psi^\prime$ and the trial wavefunction. 
\item Update the walker's weight and phase as $W \to W \times I({\bf{x}},\overline{{\bf{x}}};\Psi)$.
\item In a constrained-path calculation, ignore the phase and apply the $\cos$ projection.
\item Repeat steps 2 to 6 for all walkers. This forms one step of the random walk.
\item Periodically perform the branching and population control procedure.
\item Periodically re-orthonormalize the orbitals of the walkers.
\item Repeat this process until an adequate number of measurements has been collected.
\end{enumerate}

\section{Recent methodological advances}

The development and application of AFQMC in molecular systems is an active area of research.
Methods are maturing so that applications should start to increase rapidly. 
Recent methodological advances include the computation of observables, correlation functions, 
and geometry optimization; strategies for computing energy differences via correlated sampling; 
incorporation of functionalities of other quantum chemistry and electronic structure methods to 
extend the reach of AFQMC and improve computational efficiency.
Concurrently, many other algorithmic advances have been made, including the computation of 
excitations and imaginary-time correlation \cite{Motta_JCP_2014,Vitali_PRB_2016}, the treatment 
of spin-orbit coupling \cite{Rosenberg_arxiv_2017}, the use of self-consistent constraints 
\cite{Qin_PRB_2016}, etc, which to date have mostly been applied in lattice models of correlated 
electrons and which will not be covered here due to space limitations.

\subsection{Computing ground-state properties: the backpropagation algorithm}

The importance sampling transformation provides a stochastic representation of the ground-state 
wavefunction, and gives the possibility of computing 
the mixed estimator of an observable $\hat{A}$ as a weighted average
\begin{equation}
\mathrm{A_{mix}} = \frac{ \braket{ \Psi_T | \hat{A} | \Psi_0} }{ \braket{ \Psi_T | \Psi_0} }
\simeq 
\sum_{k} \frac{W_{n,k}}{\sum_{k} W_{n,k} } \, 
\frac{ \braket{ \Psi_T | \hat{A} | \Psi_{n,k} } }{ \braket{ \Psi_T | \Psi_{n,k} } }\,.
\end{equation}
Unless $[\hat{A},\hat{H}]=0$, the mixed estimator of $\hat{A}$ is biased by the trial wavefunction 
$\Psi_T$ used for importance sampling. In order to remove this bias, 
a back-propagation (BP) technique was proposed
\cite{Zhang_PRB55_1997,Purwanto_PRE70_2004,Motta_backpropagation_2017} 
in the framework of AFQMC. The starting point of the BP algorithm is the observation that
\begin{equation}
\label{eq:bp1}
\mathrm{A_{bp}} \equiv
\frac
{ 
\braket{ \Psi_T | e^{-m \Delta\tau \hat{H}} \hat{A} e^{-n \Delta\tau \hat{H}} | \Psi_I} 
}{ 
\braket{ \Psi_T | e^{- (m+n) \Delta\tau \hat{H}} | \Psi_I} 
}
\simeq
\frac{ \langle \Psi_0 | \hat{A} | \Psi_0 \rangle }{ \langle \Psi_0 | \Psi_0 \rangle }
\end{equation}
for large $n$ and $m$. Inserting \eqref{eq:hs2} into \eqref{eq:bp1} yields 
\cite{Purwanto_PRE70_2004,Motta_backpropagation_2017}
\begin{equation}
\mathrm{A_{bp}} = 
\frac
{ 
\int d\vett{X} \, p(\vett{X}) \braket{ \Phi_{m}(\vett{X}) | \hat{A} | \Psi_n(\vett{X}) }
}{ 
\int d\vett{X} \, p(\vett{X}) \braket{ \Phi_{m}(\vett{X}) |      \Psi_n(\vett{X}) }
}\,,
\end{equation}
where we have introduced a shorthand for the collection of auxiliary fields along a path segment:
$\vett{X} = (\vett{x}_{n+m-1} \dots \vett{x}_0)$, and for the ``back-propagated'' SD:
\begin{equation}
\label{eq:Bprod-bp}
\ket{ \Phi_{m}(\vett{X}) } \equiv \hat{B}(\vett{x}_{n})^\dagger \dots \hat{B}(\vett{x}_{n+m-1})^\dagger \ket{ \Psi_T }\,.
\end{equation}
As before, we will use the abbreviation $\Phi_m$ to denote $\Phi_{m}(\vett{X})$. Performing a shift 
of the auxiliary-field path $\vett{X}$ and applying the same similarity transformations discussed 
earlier (Importance sampling transformation) on the entire path yield
\begin{equation}
\label{eq:bp_formula}
\mathrm{A_{bp}} = 
\frac
{ 
\int d\vett{X} \, p(\vett{X}) \, W_{n+m}(\vett{X},\overline{\vett{X}}) 
\, 
\frac{ \braket{ \Phi_m | \hat{A} | \Psi_n } }
       { \braket{ \Phi_m |      \Psi_n } } 
}{ 
\int d\vett{X} \, p(\vett{X}) \, W_{n+m}(\vett{X},\overline{\vett{X}}) 
}\,.
\end{equation}
In other words, $\mathrm{A}_{bp}$ is estimated as the following weighted average, using ``retarded
weights'' and the back-propagated SDs:
\begin{equation}
\label{eq:bp}
\mathrm{A_{bp}} \simeq
\sum_{w} \frac{W_{n+m,w}}{\sum_{w} W_{n+m,w} } \, 
\frac{ \braket{ \Phi_{m,w} | \hat{A} | \Psi_{n,w} } }{ \braket{ \Psi_{m,w} | \Psi_{n,w} } } \,,
\end{equation}
with
\begin{equation}
\label{eq:bp_dets}
\begin{split}
\ket{ \Phi_{m,w} } &= 
\hat{B}^\dagger\left((\vett{x}-\overline{\vett{x}})_{n,w}\right) 
\dots 
\hat{B}^\dagger\left((\vett{x}-\overline{\vett{x}})_{n+m-1,w}\right) 
\ket{ \Psi_T }\,, \\
\ket{ \Psi_{n,w} } &= 
\hat{B}\left((\vett{x}-\overline{\vett{x}})_{n-1,w}\right) 
\dots 
\hat{B}\left((\vett{x}-\overline{\vett{x}})_{0,w}\right)  \ket{ \Psi_I }\,. \\
\end{split}
\end{equation}

In the absence of the constraints for controlling the phase problem,
Eq.~\eqref{eq:bp} approaches the exact expectation value as the 
number $m$ of BP steps is increased.
With the phaseless constraint, the BP results will incur a bias.
The most straightforward approach, which we will refer to as BP-PhL, 
applies Eq.~\eqref{eq:bp} as is \cite{Purwanto_PRE70_2004}. In 
molecular systems, this was found to cause substantial biases in some cases, and  
an improved method was proposed \cite{Motta_backpropagation_2017}.
The  new approach, referred 
to as BP-PRes (for path restoration), 
significantly reduces bias by modifications to the weights used 
in BP to partially mitigate the constraints, as described next.
  
\begin{figure}[ht!]
\begin{center}
\includegraphics[width=0.45\textwidth]{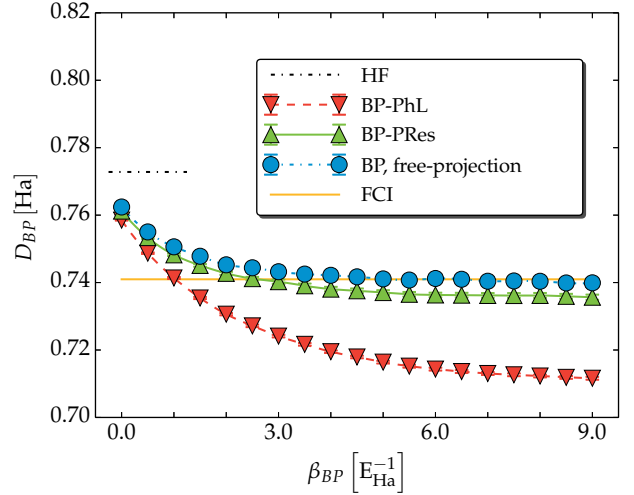}
\caption{(color online) 
Illustration of the back-propagation algorithm: evolution of the dipole moment is shown
versus back-propagation time using phaseless formalism (BP-PhL), path restoration 
(BP-PRes) and free-projection (BP, free-projection) to obtain results of increasing quality.
Adapted from \cite{Motta_backpropagation_2017}. The system is \chem{NH_3} (STO-3G 
level, trigonal pyramid geometry, $R_{\chem{NH}} = 1.07$ \AA, $\theta_{\chem{HNH}} = 
100.08^o$)
} 
\label{fig:nh3}
\end{center}
\end{figure}

The full importance function $I(\vett{x},\overline{\vett{x}}, \Psi )$, as discussed earlier, 
is equal to the exponential of the local energy, $e^{ \Delta\tau \left( \mathcal{E}_0 - 
\mathcal{E}_{loc}(\Psi) \right) }$.
The phaseless approximation removes the phase $\mbox{Im} \,\mathcal{E}_{loc}(\Psi)$, 
and multiplies the weight by the cosine projection. 
The approximation is based on a gauge condition for the \emph{forward direction}, to 
break the rotational symmetry in the imaginary-time projection along the path from steps 
$0$ to $(n+m)$. However, the constraint in AFQMC breaks (imaginary-)time symmetry 
\cite{Zhang_PRL_1999}, and the phaseless constraint applied in the forward direction 
is not optimized for the back-propagated path. 
The broken symmetry can be partially restored in the paths that survive from step $n$ 
to step $(n+m)$ in the forward direction, by retrieving the phases and undoing the projection, 
i.e., by replacing $W_{n+m,w}$ with 
\begin{equation}
W'_{n+m,w} \equiv W_{n+m,w} \prod_{k=n+1}^{n+m} 
\frac
{ e^{- \Delta\tau \, i \, \textrm{Im} \,\mathcal{E}_{loc}(\Psi_{k,w}) } 
}{ 
\max\left(0,\cos(\Delta \theta_{k,w}) \right) 
}
\end{equation}
in  Eq.~\eqref{eq:bp_formula}. 
(It is worth emphasizing that this substitution is only in the BP estimator. After BP is performed, 
the random walk resumes in the forward direction at time-step $(n+m)$, 
with the original weight $W_{n+m,w}$.)
The effectiveness of this procedure is illustrated in Fig.~\ref{fig:nh3}.

\subsection{Frozen-core, embedding, and downfolding}

Given the general form of the Hamiltonian in Eq.~\eqref{eq:bo_ham2}, the treatment of core 
electrons is straightforward. An effective core potential (ECP) can be implemented in the same 
manner as in standard quantum chemistry methods, using a modified Hamiltonian and the 
appropriate basis sets.
(Similarly, norm-conserving pseudopotentials have been applied with plane-wave basis 
\cite{Zhang_PRL90_2003,Ma_PRB_2016}.)

The frozen-core approximation is also straightforward to introduce in AFQMC
\cite{Purwanto_JCTC9_2013,Ma_PRL114_2015}. 
We proceed from the following separability approximation for a partition of the many-body ground 
state into ``active" (A) and ``inactive'' (I) spaces
\begin{equation}
\label{eq:fc}
\begin{split}
|\Psi_{gs}\rangle&= \left( \prod_{i=1}^{N_I} 
\crt{\varphi_i \uparrow} \crt{\varphi_i \downarrow} \right) |\Psi_A\rangle\,,
\end{split}
\end{equation}
where $\Psi_A$ is a wavefunction describing $(N_\sigma - N_I)$ spin-$\sigma$ electrons 
($\sigma=\uparrow$, $\downarrow$) in the Fock space spanned by the active orbitals
$A=\left\{ \varphi_{p}\right\} _{p=N_I+1}^{M}$.
The first $N_I$ electrons of each spin occupy inactive orbitals, typically defined at a lower 
level of theory (HF or DFT, for example). 
These orbitals are constrained to remain doubly-occupied and frozen. 
Active and inactive orbitals are mutually orthogonal and individually orthonormal; they span 
different parts of the one-electron Hilbert space.
To determine the ground-state wavefunction of the system within the FC approximation, we 
thus only need to simulate an effective active space Hamiltonian $\hat{H}_{A}$ with AFQMC
\cite{Purwanto_JCTC9_2013}
\begin{equation}
\label{eq:ham_fc}
 \hat{H}_A =H'_{A,0} + \hat{H'}_{A,1} + \hat{H}_{A,2}\,,
\end{equation}
where $ \hat{H}_{A,2}$ is the usual  two-body interaction projected on the active space, but 
$H'_{A,0}$ contains $H_{A,0}$ plus an extra constant from the core electrons,
while $\hat{H'}_{A,1}$ contains terms representing core-valence (I-A) interactions, similar to 
a pseudopential, in addition to the one-body operator projected on the active space, 
$ \hat{H}_{A,1}$. Additional details on Eq.~\eqref{eq:ham_fc} are given in the Appendix. 

\begin{figure*}[ht!]
\centering
\includegraphics[width=0.44\textwidth]{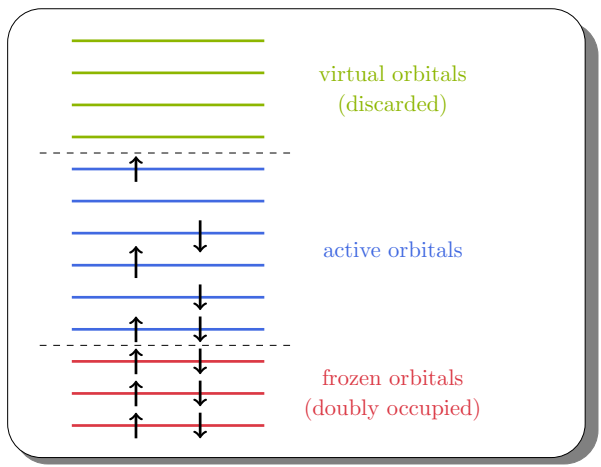}
\hspace{0.5cm}
\includegraphics[width=0.44\textwidth]{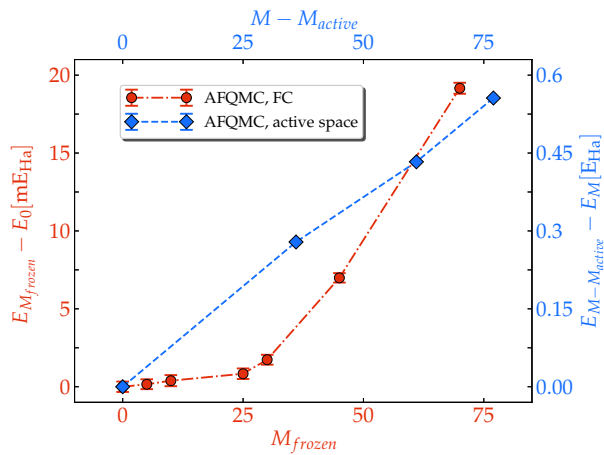}
\caption{(color online) Left: Illustration of the downfolding approach for solids 
\cite{Ma_PRL114_2015} using the idea of active space in a molecular system.
A set of molecular orbitals are first obtained, e.g., with DFT or ROHF.
A certain number ($M_{\rm frozen}=3$ in the figure) of orbitals can be frozen
and then a certain number of high-energy orbitals ($M-M_{\rm active} = 4$) 
can be discarded. Matrix elements  for the effective Hamiltonian in the remaining 
active space are obtained for use in AFQMC.
Right: Computed total energy in the Kr atom relative to the reference value.
Red points: versus  the size of the frozen core  (cc-pVDZ basis); 
blue diamonds: versus the size of truncation to the active space (cc-pV5Z  basis).
In both cases RHF orbitals are used as basis of the one-electron Hilbert space. 
} 
\label{fig:fc}
\end{figure*}

The same approach also allows a completely different application, to embed a local 
region of the system in an environment frozen at the independent-electron level. An example 
of this was given in a study of Co adsorption on graphene, in which AFQMC treated 
the Co atom and the closest C atoms in graphene, while outer rings of graphene were 
frozen in DFT orbitals \cite{Virgus_PRB86_2012}.

Another idea that proved very effective for treating periodic solids is that of downfolding,
which has provided excellent accuracy while significantly reducing 
the computational cost \cite{Ma_PRL114_2015}.
DFT calculations involving large plane-wave bases are first performed, yielding a set 
of $M_{\rm PW}$ orbitals. 
The downfolding technique consists of retaining the first $M$
low-energy DFT orbitals,  $\left\{ \varphi_{p}\right\} _{p=1}^{M}$ (with $M\ll M_{\rm PW}$),
as one-particle basis set for AFQMC. 
The one-body and two-body matrix elements can be evaluated straightforwardly 
\cite{Purwanto_JCTC9_2013} using plane-wave techniques.  
For an extended system with inversion symmetry, any twist ${\mathbf k}$ can be applied 
to the supercell, and the resulting matrix elements are real-valued \cite{Ma_PRL114_2015},
and can therefore be fed into AFQMC or any quantum chemistry code. Typically this was 
applied in conjunction with a frozen core approximation in which core orbitals are frozen 
at the DFT level. This has allowed more accurate calculations compared to conventional 
pseudopotentials, for example for systems under high pressure, where the core orbitals are 
obtained in the solid environment with DFT before they are frozen \cite{Ma_PRL114_2015}.  
 
One way to illustrate the downfolding approach is to use the idea of active space in quantum 
chemistry, as in Fig.~\ref{fig:fc}. To reduce the degrees of freedom in a many-body system,
one can  focus on the low-energy, physically relevant sector of the Hilbert space.
First, an independent-electron calculation such as ROHF or restricted DFT (or even a low-level 
correlated calculation with modest computational cost) is performed, yielding a set of MOs. 
We can then freeze some of the lowest-energy occupied orbitals and discard some of 
the highest-energy virtual orbitals to create an active space. An effective Hamiltonian is then 
obtained for the active space and used to perform a standard AFQMC calculation. 
 
\subsection{Computing energy differences: the correlated sampling algorithm}

Efficient calculation of many chemical properties 
requires computing the energy difference $\Delta \mathcal{E} = 
\mathcal{E}_1 - \mathcal{E}_2$ between two closely related systems, 
termed primary and secondary. Those energies can in turn be written as expectation values 
\begin{equation}
\mathcal{E}_k = \int dx \, g_k(x) \, p_k(x) = \mathrm{E}[g_k(X_k)]
\end{equation}
of random variables $g_k(X_k)$, with $X_k$ distributed according to $p_k$. In uncorrelated 
calculations, $\Delta \mathcal{E} = \mathrm{E}[g_1(X_1)-g_2(X_2)]$ is evaluated by two distinct 
MC simulations, i.e., with $X_1$ and $X_2$ statistically independent, in which case 
\begin{equation}
\mbox{var}[g_1(X_1)-g_2(X_2)] = \mbox{var}[g_1(X_1)] + \mbox{var}[g_2(X_2)] \, .
\end{equation}
With correlated sampling,  one seeks to generate $X_1$ and $X_2$ in a correlated manner, 
e.g. in terms of a random variable $Z$ (which could also be either $X_1$ or $X_2$). Then
\begin{equation}
\mathcal{E}_k 
= 
\mathrm{E}[g_k(X_k)] 
= 
\mathrm{E}[\tilde{g}_k(Z)] \,\, , \,\, \tilde{g}_k(z) 
= 
\frac{g_k(x(z)) \, p_k(x(z))}{p(z)}
\end{equation}
and the expression for the variance turns into
\begin{equation}
\mbox{var}[\tilde{g}_1-\tilde{g}_2] 
= 
\mbox{var}[\tilde{g}_1] + \mbox{var}[\tilde{g}_2] - 2 \mbox{cov}[\tilde{g}_1 \tilde{g}_2] \, ,
\end{equation}
with $ \mbox{cov}[\tilde{g}_1 \tilde{g}_2] \neq 0$. Clearly, a positive and large covariance leads 
to a significant variance reduction and a more efficient estimate of $\Delta \mathcal{E}$. 
This is achieved when $\tilde{g}_1$ and $\tilde{g}_2$ are very similar.
Within AFQMC, the energies in the reference and secondary systems ($k=1,2$) 
\begin{equation}
\mathcal{E}^{(k)}
\simeq
\frac
{ \int\prod_{l=0}^{n-1}d{\bf x}_{l}\,p({\bf x}_{l}) \, 
I^{(k)}({\bf x}_{l},\overline{{\bf x}}_{l}^{(k)};\Psi_l) \, \mathcal{E}^{(k)}_{loc}(\Psi_l) }
{ \int\prod_{l=0}^{n-1}d{\bf x}_{l}\,p({\bf x}_{l}) \, 
I^{(k)}({\bf x}_{l},\overline{{\bf x}}_{l}^{(k)};\Psi_l) }\, 
\end{equation}
are computed by sampling the auxiliary field paths $\{{\bf x}_{l}\}_{l=0}^{n-1}$ 
in the reference system ($k=1$) and using them for both systems \cite{Shee_JCTC13_2017}. 
In the secondary system (using the same basis), reweighting is performed in the estimator 
to account for $I^{(2)}/I^{(1)}$. 
Sampling in this way requires keeping the paths correlated for as long as possible 
in the imaginary-time projection.
Independent population control for the two systems should be avoided, as this operation 
would quickly destroy the desired correlations between primary and secondary system
\cite{Shee_JCTC13_2017}.
Correlated sampling has been applied to the calculation of atomic redox properties, 
and the ionization potential,  OH deprotonation and bond-dissociation energies of methanol, 
resulting in a significant improvement of computational efficiency \cite{Shee_JCTC13_2017}.

\section{Selected applications in molecular systems}

The cp-AFQMC method has been applied to a variety of molecular systems, including
first- and second- row molecules at equilibrium geometry 
\cite{AlSaidi_JCP124_2006,Suewattana_PRB75_2007,Zhang_CPC169_2005}
and along bond stretching 
\cite{AlSaidi_JCP127_2007,Purwanto_JCP128_2008,Purwanto_JCP130_2009},
transition metal oxides \cite{AlSaidi_PRB73_2006} 
and dimers \cite{Purwanto_JCP142_2015,Purwanto_JCP144_2016}, 
first- and second- row post-$d$ elements \cite{AlSaidi_JCP125_2006} 
and molecular clusters with applications to H$_2$ physiorption and spintronics 
\cite{Purwanto_JCP135_2011,Virgus_PRB86_2012}.
Recently AFQMC has also been applied to the hydrogen chain in a large-scale benchmark study
\cite{Motta_PRX_2017}, yielding state-of-the-art results.
Here we present selected applications of AFQMC to illustrate the method in molecular systems. 
There are also significant applications to solids 
\cite{Zhang_PRL90_2003,Kwee_PRL100_2008,Purwanto_PRB80_2009,Ma_PRL114_2015}, 
lattice models of correlated electrons \cite{LeBlanc_PRX5_2015, Zheng_SCI_2017} 
and ultracold atoms \cite{Shi_PRA_2015,Rosenberg_arxiv_2017} that will not be covered here.

\subsection{Benchmarks of the phaseless AFQMC}

\subsubsection{Molecular binding energies}

\begin{figure}[ht!]
\centering
\includegraphics[width=0.45\textwidth]{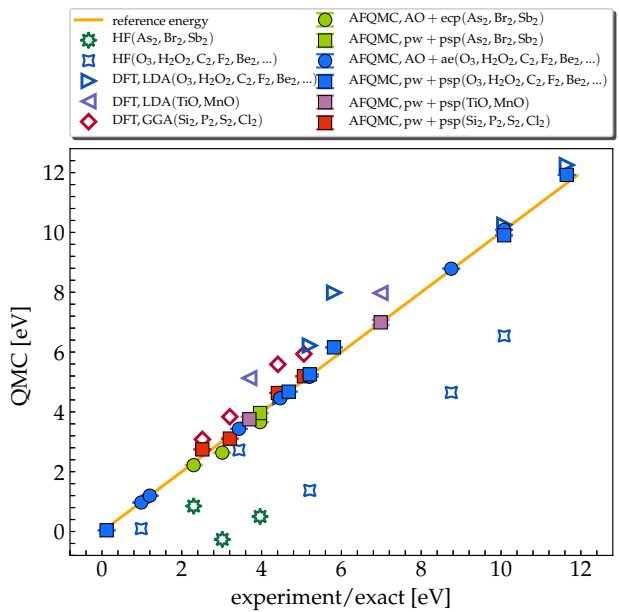}
\caption{(color online) Comparison between exact or experimental (line)
and AFQMC (filled symbols) binding energies for several molecules.
AFQMC calculations with plane-waves and pseudopotentials (pw+psp), Gaussian atomic orbitals (AO+ae, 
AO+ecp for all-electron and effective-core potential calculations) are marked with filled squares and circles 
respectively. Green, blue, purple and red symbols refer to first- and second- row post-d elements 
\cite{AlSaidi_JCP125_2006}, 
sp-bonded materials 
\cite{Zhang_PRL90_2003,AlSaidi_JCP124_2006,AlSaidi_PRB73_2006,
AlSaidi_JCP126_2007,Purwanto_JCP128_2008,Purwanto_JCP130_2009},
transition metal oxides \cite{AlSaidi_PRB73_2006} and dimers respectively \cite{Zhang_CPC169_2005}.
The AFQMC takes as its initial and trial wave function a single determinant from HF or DFT; 
the corresponding results from these independent-electron calculations are also shown (HF: empty stars, 
DFT-LDA: empty triangles, DFT-GGA: empty diamonds). 
} 
\label{fig:be}
\end{figure}

Early applications of AFQMC benchmarked the methodology against exact results and experimental 
data for a variety of systems, from $sp$-bonded materials 
\cite{Zhang_PRL90_2003,AlSaidi_JCP124_2006,AlSaidi_PRB73_2006,AlSaidi_JCP126_2007,
Purwanto_JCP128_2008,Purwanto_JCP130_2009} to first- and second- row post-$d$ elements 
\cite{AlSaidi_JCP125_2006}, 
transition metal oxides 
\cite{AlSaidi_PRB73_2006} 
and dimers 
\cite{Zhang_CPC169_2005}.
Some of the binding energies computed in these systems are shown in Fig.~\ref{fig:be} in comparison 
with reference experimental or exact results. 
The AFQMC calculations include three different types: 
all-electron (ae) using AO basis as is typical in quantum chemistry; 
AO basis using small-core effective-core potentials (ecp);
plane-wave (pw) and pseudopotentials (psp) as is typical in solid-state calculations,
which place the molecule (or the atom) in a large periodic supercell. 
These are indicated by different symbols. 
As is seen, AO+ae, AO+ecp and PW+psp calculations all lead to excellent agreement with the experimental 
result (zero-point energy removed). 
(In a few of the cases, an exact result for a finite AO basis is shown in lieu of experiment, and compared 
with the corresponding AFQMC calculation using the same basis.)
 
The initial and trial wave function used in these calculations were taken to be a single determinant, 
either HF or DFT. The results from the corresponding independent-electron calculations that generated 
the $|\Psi_T\rangle$ are also shown.
In a few cases, two different types of trial wavefunctions are tested, and two sets of symbols are seen at 
the same horizontal coordinate. It is seen that the AFQMC results from two different $|\Psi_T\rangle$'s  
are in good agreement with each other and with experiment,
showing the typical  weak reliance of  AFQMC on the trial wavefunction.
By feeding in a DFT (or HF) solution, a reliable correction is obtained to the total energy.
These and other results demonstrate AFQMC's potential as a general post-DFT method for  molecular systems.

\subsubsection{Bond breaking}

The bond stretching process exhibits a varying nature of chemical bonding and electron correlation: in the compressed 
regime (small internuclear distances) the ground state is shaped by dynamical correlations, accurately captured by 
single-reference methods like coupled cluster; in the dissociation regime (large internuclear distances), due to 
quasi-degeneracies, there can be more than one important electron configuration, and a single Slater determinant 
cannot often adequately describe the system.

AFQMC has been applied to study bond stretching and breaking in several chemical systems, from well-studied 
molecules like H$_2$O \cite{AlSaidi_JCP124_2006}, BH and N$_2$ \cite{AlSaidi_JCP127_2007} and F$_2$ \cite{Purwanto_JCP128_2008} 
to the more challenging case of hydrogen chains \cite{AlSaidi_JCP127_2007,Motta_PRX_2017}.
In all cases, with a single Slater determinant from unrestricted HF (UHF)  or a very
compact multideterminant trial wavefunction, the ph-AFQMC method generally gives better overall accuracy and a 
more uniform behavior than CCSD(T). Chemical accuracy can be achieved  with  ph-AFQM
in the description of  the potential-energy curve across 
bond stretching  using a modest $|\Psi_T\rangle$.

\begin{figure*}[t!]
\centering
\includegraphics[width=0.45\textwidth]{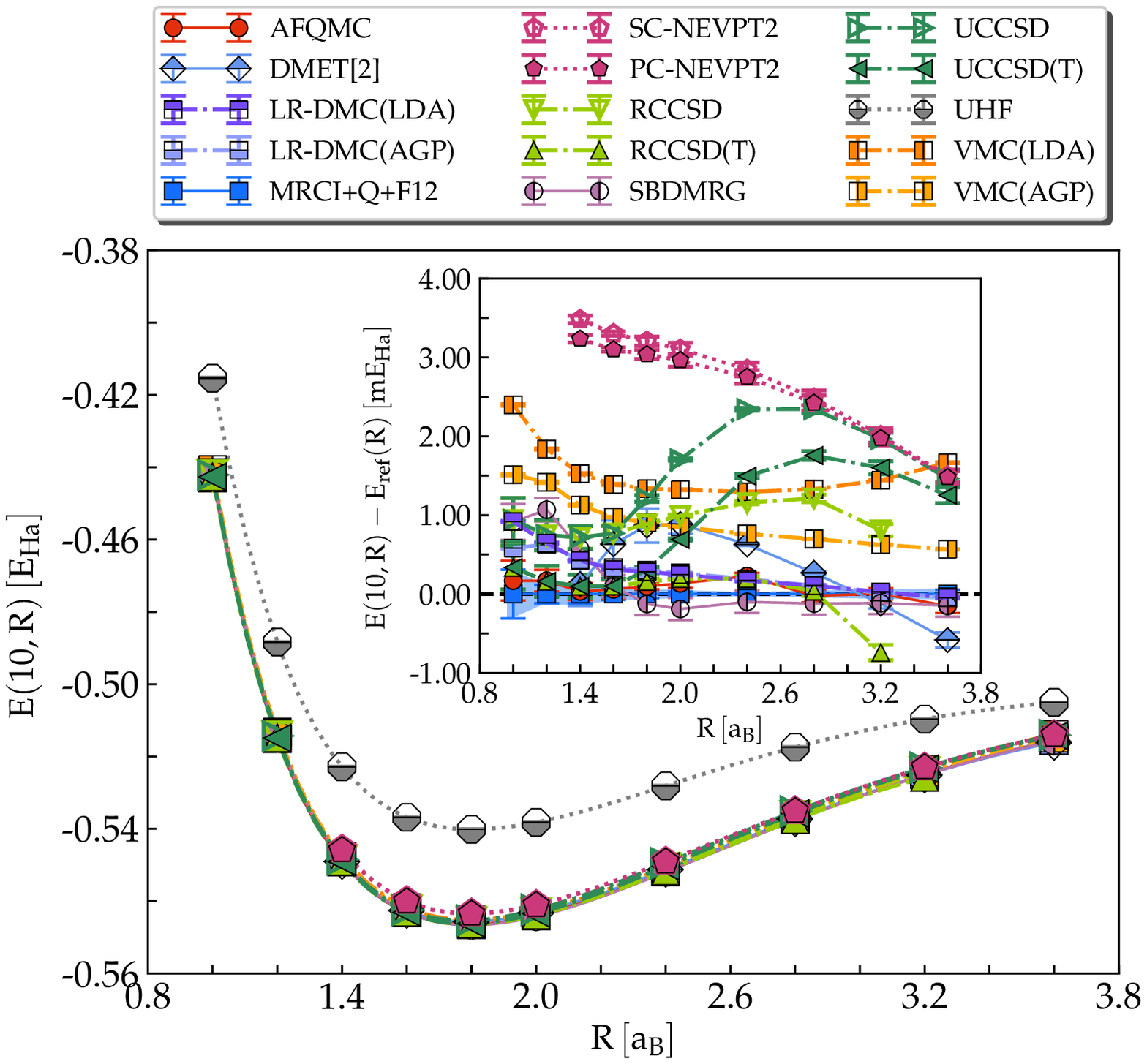}
\includegraphics[width=0.45\textwidth]{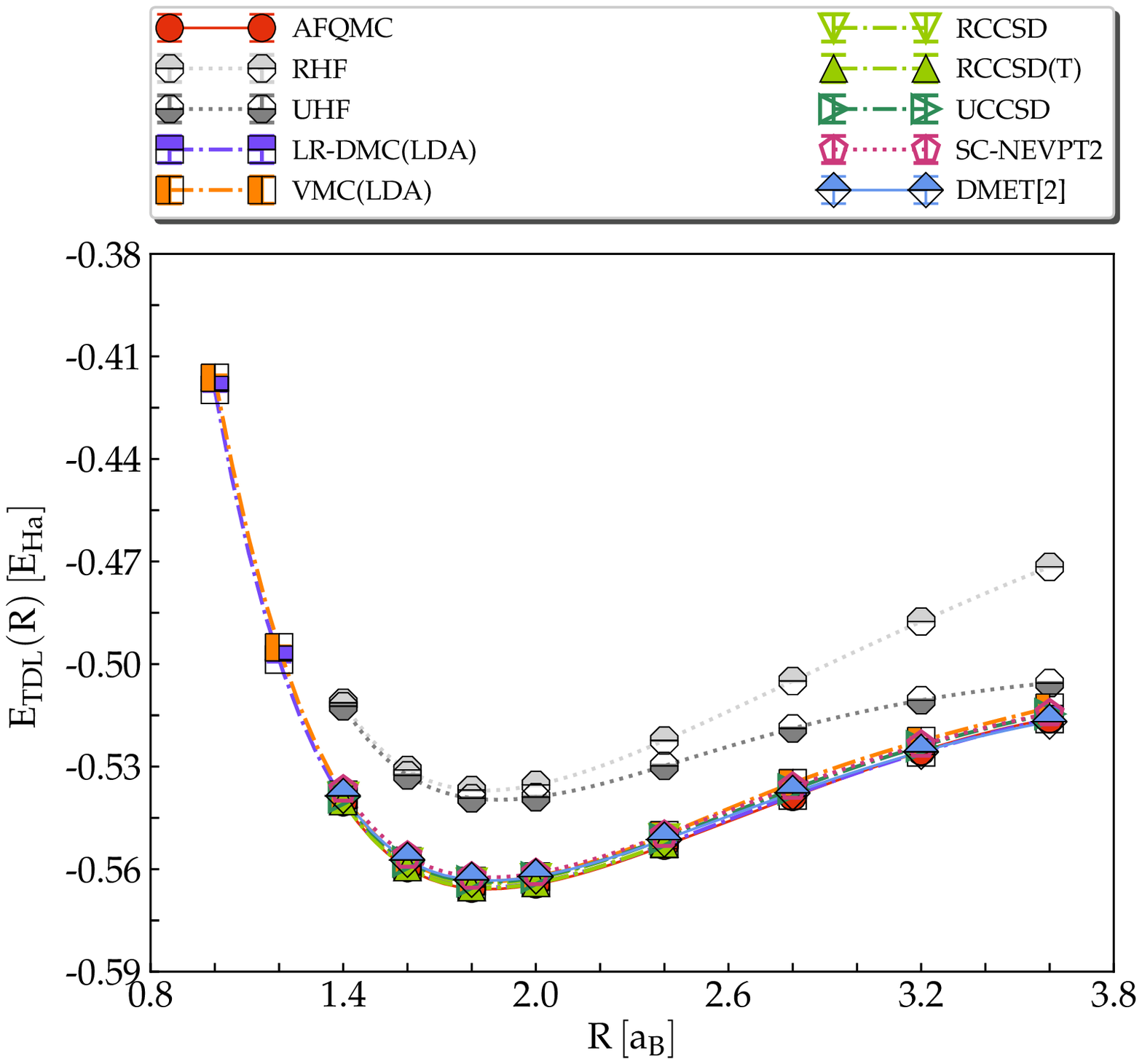}
\caption{
Left: Energy per particle of H$_{10}$ in the CBS limit, versus the distance $R$ between two 
consecutive H atoms, from a recent benchmark study \cite{Motta_PRX_2017}. Inset: comparison of the \chem{H_{10}} energies per particle, 
using MRCI+Q+F12 results at CBS as reference \cite{Motta_PRX_2017}.
The excellent agreement between AFQMC and MRCI+Q, MRCI+Q+F12 confirms the uniform and overall high accuracy 
of the methodology. Right: equation of state of the H chain,  extrapolated to the joint 
CBS and thermodynamic limit \cite{Motta_PRX_2017}.
} \label{fig:hchain1}
\end{figure*}

Bond stretching in a linear chain of equally spaced hydrogen atoms was recently benchmarked \cite{Motta_PRX_2017} with a variety of
state-of-the-art methods, including AFQMC. As illustrated in Fig.~\ref{fig:hchain1}, the method is capable of delivering 
remarkably accurate equations of state in the challenging and chemically and physically relevant complete basis 
set and thermodynamic limits.

\subsubsection{Eliminating spin contamination and restoring spatial symmetries}

Imposition of symmetry properties results in a significant increase in the efficiency and accuracy of AFQMC, and often 
can be achieved with very modest computational overhead. In this Section, we give a brief discussion of symmetry, 
focusing on the elimination of spin contamination 
\cite{Purwanto_JCP128_2008} and the restoration of spatial symmetries \cite{Shi_PRB88_2013,Shi_PRB89_2014}.

Often the system has symmetries, i.e., operators that commute with the Hamiltonian. Symmetries of the Hamiltonian 
are important constraints for formulating physical theories and models; they are also instruments for achieving accurate 
computational treatments. In particular, it is desirable that approximate eigenfunctions of $\hat{H}$ are eigenstates of the
Hamiltonian symmetries with suitable eigenvalues.

One important symmetry in molecular systems is total spin. The total spin-squared, $\hat{\bf{S}}^2$, commutes with 
its $z$ component $\hat{S}_z$ and $\hat{H}$. 
ROHF determinants are eigenfunctions of $\hat{\bf{S}}^2$ with eigenvalue $\lambda = \frac{N_\uparrow-N_\downarrow}{2} 
\left( \frac{N_\uparrow-N_\downarrow}{2} + 1 \right)$ (assuming $N_\uparrow \geq N_\downarrow$) \cite{Szabo_book_1989}, 
while UHF determinants in general are not. 
Approximate wave-functions that are (are not) eigenfunctions of $\hat{\bf{S}}^2$ are called spin pure (contaminated).

\begin{figure}
\centering
\includegraphics[width=0.45\textwidth]{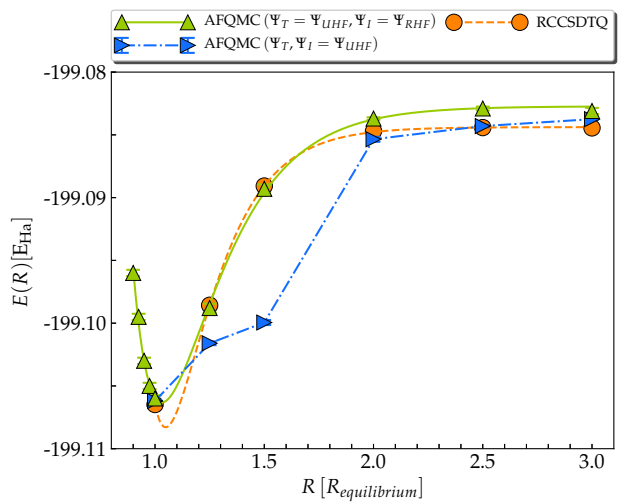}
\caption{
Elimination of spin contamination in AFQMC calculations, adapted from \cite{Purwanto_JCP128_2008}. 
The trial wave function $|\Psi_T\rangle$ is UHF.
Using ROHF as $|\Psi_I\rangle$ to initialize the population restores spin symmetry in the walkers.
The AFQMC energies have an overall better accuracy and a more systematic behavior, without increasing the 
cost of the simulation. 
The system is F$_2$ at cc-pVDZ level, and RCCSDTQ results are from \cite{Musial_JCP122_2005}.
} \label{fig:F2} 
\end{figure}

From an unrestricted initial wave-function $|\Psi_I\rangle$ (e.g., UHF), the AFQMC projection will restore spin symmetry
\emph{statistically} and  lead to a spin-pure ground state, if no approximation is present. 
Each random walker individually can remain spin-contaminated, however. 
If a spin-contaminated trial wavefunction $|\Psi_T\rangle$ is used in the phaseless constraint, a contaminated spin 
component can survive, causing larger systematic errors in the calculation. 
This is illustrated in Fig.~\ref{fig:F2}, using the F$_2$ molecule as a test case.

One way to eliminate the spin contamination is to ensure that the population of random walkers consists of spin-pure 
Slater determinants. To achieve this goal,  it is sufficient to use an ROHF-type wave function as $|\Psi_I\rangle$ to 
initialize the population, since the HS transformation discussed earlier, which takes the form of a Hartree or charge 
decomposition, preserves spin symmetry: $[\hat{B}({\bf{x}}),\hat{\bf{S}}^2]=0$. In this case, since each walker in the 
population remains spin-pure along the imaginary-time projection, the spin-contaminated component in the trial 
wave function $|\Psi_T\rangle$ can be projected out while still retaining some of the  benefit of the UHF in the 
phaseless approximation.
In addition to removing spurious spin contamination effects in the calculated AFQMC energy, spin-projection can 
sometimes also reduce the imaginary time needed for equilibration by excluding certain excitations \cite{Purwanto_JCP128_2008}.

To impose spatial symmetries $\hat{R}$ is more subtle, because generally $[\hat{B}({\bf{x}}),\hat{R}]\ne 0$.
Restoring symmetry in the trial wave function can help project the walkers and is often helpful \cite{Shi_PRB88_2013}.
Broken spatial symmetries are often encountered in open-shell systems (e.g., in atoms) at the HF level. 
These can be restored performing a complete active space expansion in the subspace of the open shell, 
or relying on the symmetry restoration technique \cite{Shi_PRB89_2014,Purwanto_JCP144_2016}.

\subsubsection{Excited states calculations}

The ability to calculate electronic excited states of molecules and extended systems is necessary to predict
key phenomena and properties of technologically important systems. 
It remains one of the leading challenges in electronic structure theory.

Calculating the energy of the lowest excited states belonging to an irreducible representation of the symmetry 
group of $\hat{H}$ different from that of the ground state is straightforward.
It is sufficient to choose a trial wave-function $\Psi_T^*$ with the symmetry of the desired excited state and, 
naturally, an initial population with $\langle \Psi^*_T | \Psi^*_I \rangle \neq 0$. Under these conditions, 
$\mathcal{E}^*(n\Delta\tau) 
= \frac
{ \langle \Psi^*_T | \hat{H} e^{-n \Delta\tau \hat{H}} | \Psi_I^* \rangle }
{ \langle \Psi^*_T | e^{-n\Delta\tau \hat{H}} | \Psi_I^*  \rangle }
$
converges towards the energy of the lowest excited state of the targeted irreducible representation. Although 
it is often impossible for a single-reference $|\Psi_T\rangle$, multi-reference wave-functions can to a very 
high degree satisfy the desired symmetry requirement.

\begin{figure}[t!]
\centering
\includegraphics[width=0.45\textwidth]{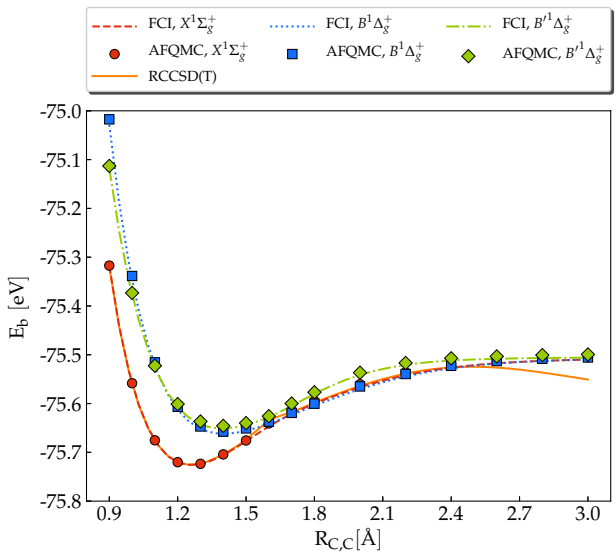}
\caption{
Equation of state of C$_2$ at 6-31G* level from AFQMC (red points, blue squares, green diamonds) using 
a truncated CASSCF(8,16) trial wavefunction, FCI (red dashed line, green dotted line, blue dot-dashed line) 
and RCCSD(T) (solid orange line).
} \label{fig:c2} 
\end{figure}

Computing energies of excited states belonging to the same irreducible representation of the ground-state is 
more challenging. It is intrinsically difficult to maintain orthogonality with the lower-lying states
when the targeted many-body excited state is being represented \emph{stochastically} in an imaginary-time projection 
method. If a trial wavefunction of sufficient quality is available, $\mathcal{E}^*(n\Delta\tau)$ can converge to the 
desired results.
Application to the C$_2$ molecule \cite{Purwanto_JCP130_2009}, sketched in Fig.~\ref{fig:c2}, indicate that 
multi-reference trial wave-functions with a modest number of configurations, e.g. suitably truncated CASSCF 
states, are often adequate for this task.
Collapse onto the ground state can be further prevented by imposing additional orthogonality constraints 
\cite{Ma_NJP15_2013}. This technique has been applied to the calculation of optical gaps in bulk solids 
\cite{Ma_NJP15_2013}.
Recent advances in computing imaginary-time Green's functions can provide another route for determining 
excitation gaps \cite{Vitali_PRB_2016}.

\subsubsection{Transition metal dimers}

The study of transition metals presents an outstanding challenge for many-body electronic structure methods, 
because of strong static and dynamic electronic correlations.
The complicated nature of binding encountered in transition metal dimers is emblematic of the delicate 
balance between competing tendencies found in many strongly correlated materials. The relatively small size 
of these systems makes them more amenable to systematic and rigorous theoretical studies.
Group VIB dimers are interesting, because the atom fragments are in the high-spin state ($^7$S), and they 
form a closed shell (1$\Sigma$) configuration in the molecular ground state, resulting in a many-body spectrum 
with many nearly degenerate states and strong electronic correlation effects.

Recently, AFQMC has been successfully applied to investigate the equation of state of two such dimers, 
Mo$_2$ \cite{Purwanto_JCP144_2016} and Cr$_2$  \cite{Purwanto_JCP142_2015}.
For the Mo$_2$ molecule, the best quantum chemistry calculations have given widely varying predictions.
Cr$_2$ is a well-known challenge, featuring a formal sextuple bond, with a weak binding energy 
($\sim$ 1.5 eV), a short equilibrium bond length ($\sim$ 1.7 \AA) \cite{Bondybey_CPL94_1983}, and an exotic 
shoulder structure in the intermediate to large $R$ region \cite{Casey_JPC97_1996}.

\begin{figure*}[t!]
\centering
\includegraphics[width=0.8\textwidth]{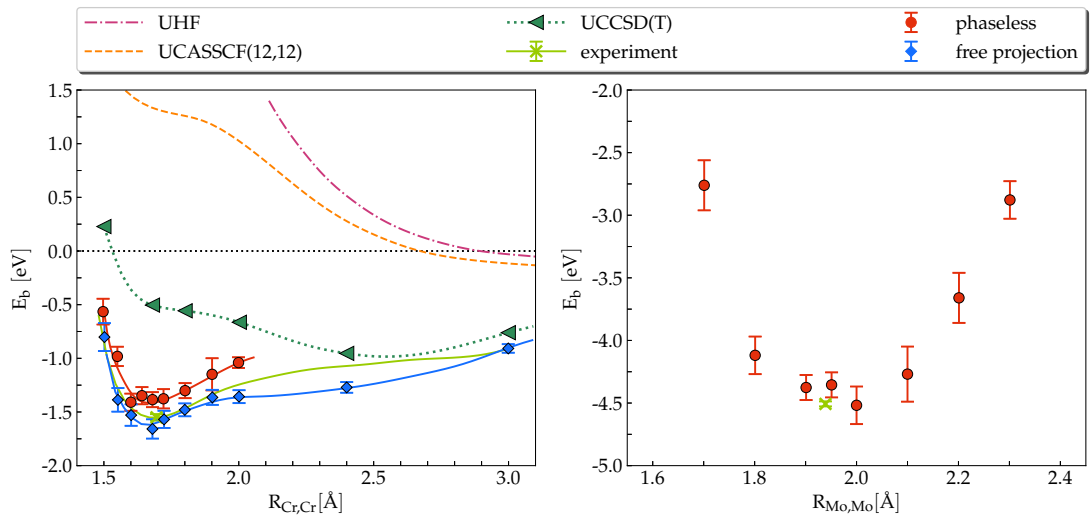}
\caption{
Left: CBS-extrapolated equation of state of Cr$_2$ from ph-AFQMC, free-projection AFQMC 
(red circles, blue diamonds)  \cite{Purwanto_JCP142_2015} compared to experiment
\cite{Casey_JPC97_1996,Simard_JCP108_1998} (solid green line), UHF, UCASSCF(12,12) and UCCSD(T).
Right: CBS-extrapolated equation of state of Mo$_2$ from ph-AFQMC  (red circles) \cite{Purwanto_JCP144_2016} 
compared to experiment \cite{Kraus_PCC4_2001}
} \label{fig:transition} 
\end{figure*}

In both cases, AFQMC has provided one of the most accurate theoretical results of the potential energy 
curve \cite{Casey_JPC97_1996,Simard_JCP108_1998}, as illustrated in Fig.~\ref{fig:transition}.
Strong static correlation present in both molecules have required the use of multi-determinant trial wave-functions, 
and selected benchmark calculations have been performed within the exact free-projection scheme, to establish 
the accuracy of constrained calculations. Results are extrapolated to the CBS limit with standard methods.

\section{Outlook and conclusions}

Despite the foundational difficulty of solving the many-electron Schr\"odinger equation,
our ability for accurate computations is steadily increasing thanks to the development 
of new numerical methods 
\cite{White_PRL69_1992,Bartlett_RMP79_2007,Zhang_PRL90_2003,Knizia_PRL109_2012,
Kurashige_NAT5_2013,Booth_NAT493_2013,Yang_SCI345_2014,LeBlanc_PRX5_2015,Motta_PRX_2017}.
With the rapid growth in computing power, computational methods will be increasingly relied upon 
for insight into properties of materials. The continued development of methods and algorithms for 
interacting quantum many-body systems will be a key in the materials genome initiative.

In this review, we have provided an expository introduction to the AFQMC method, focusing on applications 
to molecular systems. 
Among the approaches to approximately solve the Schr\"odinger equation, AFQMC is distinguished 
by a number of positive features. It has favorable (low-power) scaling with the size of the system.
It is non-perturbative and maintains systematically 
high accuracy, as demonstrated by applications across a wide range of systems from models to 
molecules and solids, including several recent benchmark studies. 
The method is  naturally parallel and has exceptional flexibility for scaling on petascale or exascale platforms 
\cite{Esler_JP125_2008}.  Its close formal connection with HF or DFT means that it can incorporate techniques 
and algorithms from standard electronic structure methods straightforwardly, as in the implementation of frozen-core or 
pseudopotentials, spin-orbit coupling \cite{Rosenberg_arxiv_2017}, the computation of density and forces 
\cite{Motta_backpropagation_2017,Motta_geometry_2017}, and the ability to embed in or interface with 
independent-electron calculations \cite{Virgus_PRB86_2012,Zheng_PRB_2016}.
AFQMC can treat a general Hamiltonian with any one-body and two-body interactions, which offers 
many opportunities, for example the removal of chemically inactive degrees of freedom or targeting 
certain active space in the mode of model calculations, the treatment of electron-phonon interactions, 
formalism at finite-temperatues \cite{Zhang_PRL_1999}, or generalization to particle number 
non-conserving Hamiltonians or treating systems with superconducting order \cite{Shi_PRB_2017}. 
AFQMC results can be systematically improved, for example via self-consistency or by constraint-release.

Owing to these features, AFQMC is  a promising method of choice in the computational study of 
material systems. In many of the directions mentioned above, only the surface has been scratched 
in the development and application of AFQMC. 
There is undoubtedly many other directions that have not even been discussed here.
With continued and increased effort in the development of algorithms and codes \cite{Nguyen_CPC_2014},
broader application of AFQMC will be greatly accelerated. 
We hope that this review will help spur more effort in the diffusion, application and development of AFQMC.

\section{Acknowledgements}

We are indebted to many colleagues for helpful interactions and collaborations which made this work possible. In
particular we would like to thank H.~Krakauer, F.~Ma, W.~Purwanto, W.~A.~Al-Saidi, J.~Shee, B.~Eskridge and
H.~Shi.
We gratefully acknowledge the NSF (Grant no. DMR- 1409510) and the Simons Foundation for support. We also
thank the Oak Ridge Leadership Computing Facilities, the NSF XSEDE, and W\&M High-Performance Computing
Facilities for computational resources.

\appendix

\section{Additional details on formalism}

In this section we prove Theorem \ref{thm:slater}, and the Hubbard-Stratonovich transformation formula
\eqref{eq:hs}.
For simplicity we consider an orthonormal basis of molecular spin-orbitals $\{ \chi_\mu \}_{\mu=1}^{2M}
= \{ \varphi_{1\uparrow} ... \varphi_{M\uparrow} \varphi_{1 \downarrow} ... \varphi_{M\downarrow} \}$
and the associated creation and destruction operators $\crt{\mu}$, $\dst{\nu}$.
This formalism simplifies and generalizes the proof, and becomes necessary when dealing with spin-orbit coupling.
Let us recall that an SD 
$|\Psi\rangle = \prod_{i=1}^N \crt{u_i} |\emptyset\rangle$ with $N_\uparrow$, $N_\downarrow$ particles  
is described by a matrix $U_{\mu i} = \langle \chi_\mu | u_i \rangle$ with the block form 
\begin{equation}
U = 
\left( 
\begin{array}{cc}
U_\uparrow & 0 \\
0 & U_\downarrow \\
\end{array}
\right)\,.
\end{equation}

\subsection{Proof of Theorem  \ref{thm:slater}}

Point 1 of Theorem \ref{thm:slater} is readily proved using the first-quantized form
\begin{equation}
\Psi({\bf x}_{1}\dots{\bf x}_{N})
=
\sum_{\sigma}\frac{(-1)^{\sigma}}{\sqrt{N!}}\prod_{l=1}^{N}\psi_{\sigma(l)}({\bf x}_{l})
\quad,
\end{equation}  
where ${\bf x}_i$ are particle coordinates and $\sigma$ denotes a permutations of $N$ objects. Thus 
\begin{equation}
\begin{split}
&\langle\Phi|\Psi\rangle = \sum_{\rho \sigma}\frac{(-1)^{\rho \sigma}}{N!} \, 
\prod_{l=1}^{N}
\int d{\bf x}_{l} \, \phi_{\rho(l)}^{*}({\bf x}_{l})\,\psi_{\sigma(l)}({\bf x}_{l})
= \\
&= \sum_{\tau} \, (-1)^{\tau} \, 
\prod_{l=1}^{N}\langle\phi_{l}^{*}|\psi_{\tau(l)}\rangle= \det\left(\Phi^{\dagger}\Psi\right)
\,.
\end{split}
\end{equation}  
To prove point 2, let us first observe that 
\begin{equation}
\hat{A}\crt{\psi} = 
\sum_{\mu\nu\sigma} A_{\mu\nu}\psi_{\sigma} \crt{\mu} \dst{\nu} \crt{\sigma} =
\sum_{\mu}(A \psi)_{\mu} \crt{\mu} + \crt{\psi} \hat{A}\,. 
\end{equation}  
Thus $\hat{A}^{n} \crt{\psi} = \sum_{s=0}^{n}\binom{n}{s} \crt{A^{s}\psi} \hat{A}^{n-s}$ is proved by 
induction over $n$, and $e^{\hat{A}} \crt{\psi} = \crt{e^{A}\psi} e^{\hat{A}}$ readily follows. Therefore
\begin{equation}
\begin{split}
   & e^{\hat{A}}|\Psi\rangle = e^{\hat{A}} \crt{\psi_{1}} \dots \crt{\psi_{N}} |\emptyset \rangle = 
\crt{e^{A}\psi_{1}} \dots \crt{e^{A}\psi_{N}} | \emptyset \rangle\,.
\end{split}
\end{equation}  
Equation \eqref{thm:thouless} and Jacobi's formula 
$\frac{d}{dt} \log \det(A_{t})= \mbox{Tr}\left[A_{t}^{-1}\frac{dA_{t}}{dt}\right]$ 
allow straightforward computations of matrix elements of one- and two-body operators: 
\begin{equation}
G^{\Phi \Psi}_{\mu\nu} = \frac{\langle\Phi| \crt{\mu} \dst{\nu} |\Psi\rangle}{\langle\Phi|\Psi\rangle}
=
\lim_{t\to0}\frac{d}{dt}\log\langle\Phi|\Psi_{t}\rangle
\end{equation}  
and
\begin{equation}
\frac{\langle\Phi| \crt{\mu} \dst{\nu} \crt{\sigma} \dst{\rho} |\Psi\rangle}{\langle\Phi|\Psi\rangle}
=
\lim_{t',t\to0}
\frac{d^{2}}{dt'\,dt}\log\langle\Phi|\Psi_{t',t}\rangle + G^{\Phi \Psi}_{\mu\nu} G^{\Phi \Psi}_{\sigma\rho}
\end{equation}  
where 
$|\Psi_{t}\rangle=e^{t \crt{\mu} \dst{\nu}} |\Psi\rangle$, 
$|\Psi_{t',t}\rangle=e^{t \crt{\mu} \dst{\nu} + t' \crt{\sigma} \dst{\rho}}|\Psi\rangle$. 
The resulting expressions are \eqref{thm:1rdm} and 
\begin{equation}
\frac{\langle\Phi| \crt{\mu} \dst{\nu} \crt{\sigma} \dst{\rho} |\Psi\rangle}{\langle\Phi|\Psi\rangle}
=
G^{\Phi\Psi}_{\mu\nu}G^{\Phi\Psi}_{\sigma\rho}
+
G^{\Phi\Psi}_{\mu\rho}\left(\delta_{\nu\sigma}-G^{\Phi\Psi}_{\nu\sigma}\right)
\end{equation}  
respectively, and the latter readily implies \eqref{thm:2rdm}. 
To prove point 5, 
let us first first observe that orthonormalizing the orbitals $u_i$ requires 
performing an invertible transformation $T$, $| u^\prime_l \rangle = \sum_{k=1}^N T_{lk} |u_k\rangle$.
Therefore
\begin{equation}
|\Psi^\prime \rangle 
= 
\crt{u^\prime_1} \dots \crt{u^\prime_N} |\emptyset \rangle 
= 
\sum_{i_1 \dots i_N=1}^N T_{1 i_1} \crt{u_{i_1}} \dots T_{N i_N} \crt{u_{i_N}} |\emptyset \rangle
\end{equation}
where the indexes $i_1 \dots i_N$ must be different from each other, due to the canonical 
anticommutation relation $a^\dagger_{u} a^\dagger_{u} = 0$, and 
thus must be a permutation of $1 \dots N$. Therefore
\begin{equation}
|\Psi^\prime \rangle 
= 
\sum_{\sigma \in S_N} 
(-1)^\sigma T_{1 \sigma(1)} \crt{\psi_1} \dots T_{N \sigma(N)} \crt{\psi_N} |\emptyset \rangle 
= 
\det(T) |\Psi\rangle\,. 
\end{equation}
For example, in the mGS procedure,  $U=U'DV$, where $D$ is diagonal and $V$ is upper triangular,
the prefactor $\det(T)$ is given by $\det(D)$   \cite{Zhang_PRB55_1997}.

\subsection{Hubbard-Stratonovich transformation \eqref{eq:hs}}

Let $\hat{A}$ be an operator on a finitely generated Fock space. Because all linear operators 
on a finite Hilbert space are bounded, the following equality holds
\begin{equation}
\int_{-\infty}^{\infty}dx\,\frac{e^{-\frac{x^{2}}{2}}}{\sqrt{2\pi}}\,e^{i\lambda x \hat{A} }
=
\sum_{n=0}^{\infty}\frac{\left(i\lambda\right)^{n}}{n!}\,J_{n}\, \hat{A}^{n}
\end{equation}  
with
\begin{equation}
J_{n}=\int_{-\infty}^{\infty}dx\,\frac{e^{-\frac{x^{2}}{2}}}{\sqrt{2\pi}}\,x^{n} = 
\left\{
\begin{array}{ll} 
0 & \mbox{$n$ odd} \\
\frac{2^{\frac{n}{2}}}{\sqrt{\pi}}\Gamma\left(\frac{n+1}{2}\right)
=
\frac{2^{-\frac{n}{2}} \, n!}{\left(\frac{n}{2}\right)!} & \mbox{$n$ even} 
\end{array}
\right.
\end{equation} 
where $\Gamma(z) = \int_0^\infty dx \, x^{z-1} \, e^{-x}$ is Euler's Gamma function.
The desired equality then follows:
 \begin{equation}
\label{eq:hstransform}
\int_{-\infty}^{\infty}dx\,\frac{e^{-\frac{x^{2}}{2}}}{\sqrt{2\pi}}\,e^{ix\lambda \hat{A}}
=
\sum_{k=0}^{\infty}\frac{1}{k!}\,\left(-\frac{\lambda^{2}}{2}\right)^{k}\,\hat{A}^{2k}
=
e^{-\frac{\lambda^{2}}{2}\hat{A}^{2}}\,.
\end{equation}  
Note that the HS transformation holds for a generic operator, not necessarily hermitian. 
Moreover, \eqref{eq:hstransform} can be written exactly as
\begin{equation}
e^{\frac{\lambda^{2}}{2} \hat{A}^{2}}
=
\int_{-\infty}^{\infty}dx\,\frac{e^{-\frac{\left(x-x_{0}\right)^{2}}{2}}}{\sqrt{2\pi}}\,e^{i(x-x_{0})\lambda \hat{A} }
 \end{equation}  
for a generic complex number $x_0 \in \mathbb{C}$. The same property is shared 
by the coefficients $J_{n}$, because the product between a Gaussian and a power 
law is a rapidly decaying holomorphic function. This additional freedom allows the 
introduction of the force bias and importance sampling in AFQMC.
  
\section{Additional algorithmic details}

\subsection{Controlling rare event fluctuations}

With the constraints, the probability distributions and the variance are well defined  in ph-AFQMC.
However, the stochastic nature of the simulation does not preclude rare events, 
which can cause large fluctuations. 
To circumvent the problem in a simulation of finite population, we apply a bound condition 
\cite{Purwanto_PRB80_2009} in the local energy:
\begin{equation}
\label{eq:cap}
\left| \mathcal{E}_{L}(\Psi)-\overline{\mathcal{E}}_{loc}\right|\leq 
\sqrt{\frac{C}{\Delta\tau}}\,,
\end{equation}
where $C$ is a constant determined by the energy units of the system (e.g., we chose $C=2$ a.u.~for molecules) 
and $\overline{\mathcal{E}}_{loc}$ is obtained by averaging the local energy during the growth phase. 
If the local energy falls outside the range \eqref{eq:cap}, it is capped at its minimum or maximum value. 
In addition, the walker weights and force biases can be bounded: $W^{(w)}\leq W_{max}$ 
and $\overline{v}_{\gamma}\leq v_{max}$. (For example, we used
$W_{max}=\max\left(10^{2},\frac{N_{w}}{10}\right)$ and $v_{max}=1$ for molecular systems with GTOs). 
Walker weights are rarely capped when the bounding scheme \eqref{eq:cap} for the local energy is active. 
Of all these precautions, the energy cap is the most effective in controlling weight fluctuations. 
Although the precise values of the
various bounds presented in this Section are somewhat empirical, the procedure 
has a well-defined behavior as $\Delta\tau\to0$: 
the bounds on $\mathcal{E}_{loc}$ and $\overline{v}_{\gamma}$ both approach $\infty$, 
hence they only affect the Trotter error at finite $\Delta\tau$ but not the final answer when $\Delta\tau$ is extrapolated to 0.

\subsection{Growth estimator of the ground-state energy}

In the projection in Eq.~\eqref{eq:togs}, $\mathcal{E}_{0}$ is the exact ground-state energy. 
In an actual calculation the unknown quantity $\mathcal{E}_{0}$ must be replaced with a tentative estimate 
or trial energy, typically $\mathcal{E}_{T}=\frac{\langle\Psi_{T}|\hat{H}|\Psi_{T}\rangle}{\langle\Psi_{T}|\Psi_{T}\rangle}$.
Deviation of the trial energy $\mathcal{E}_{T}$ from the exact result  $\mathcal{E}_{0}$ will cause the 
total weight of walkers (i.e., the population size) to, on the average,  increase or decrease
with  projection time.
The average rate of growth of the walkers' weight $R_{\Psi,{\bf {\bf x}}}=\log\frac{W[B({\bf x}-\overline{{\bf x}}) \Psi]}{W[\Psi]}$ is
\begin{equation}
\begin{split}
\label{eq:whycorsam}
&E_{\Psi,{\bf x}}\left[R_{\Psi,{\bf x}}\right] = E_{\Psi,{\bf x}}\left[\log I({\bf x},\overline{{\bf x}}; \Psi)\right]
=  \\
= \, &E_{\Psi}\left[\mathcal{E}_{loc}(\Psi)-\mathcal{E}_{T}\right]+\mathcal{O}\left(\Delta\tau^{\frac{3}{2}}\right)
=
\mathcal{E}_{0}-\mathcal{E}_{T} \,\, ,
\end{split}
\end{equation}
where the symbol $E_{\Psi,{\bf x}}$ stands for weighted average over current walker configuration $\Psi$ 
and auxiliary field ${\bf x}$. Eq.~\eqref{eq:whycorsam} offers an alternative estimator of the ground-state energy, 
the so-called growth estimator, to be averaged over many steps $n$:
\begin{equation}
\label{eq:corsam}
\mathcal{E}_{0}=\mathcal{E}_{T}
+\log\frac{\sum_{i=1}^l \sum_w W_{n+i+1}^{(w)}}{\sum_{i=1}^l \sum_w W_{n+i}^{(w)}}\,,
\end{equation}
where $l$ is an integer parameter (typically chosen to be $\mathcal{O}(10)$) that allows accumulation of
the growth ratio over $l$ steps to minimize fluctuations.
(Note that, unlike in constrained-path calculations, where the growth estimate is consistent with 
the mixed estimate \cite{Zhang_PRB55_1997},  the estimator in Eq.~\eqref{eq:corsam} 
often does not give as accurate an
estimate of the ground-state energy in ph-AFQMC because of the $\cos$ projection.)
If the population control routine allows the total number of walkers to fluctuate, we typically 
perform a relaxation phase with the trial energy, then 
adjust the trial energy with the estimator \eqref{eq:corsam} in a subsequent growth phase, 
before performing a final measurement phase.

\subsection{Frozen-core hamiltonian}

It is easy to show that two wavefunctions $\Phi,\Psi$ complying with the separability condition 
\eqref{eq:fc} have the following one-particle Green's function:
\begin{equation}
\label{eq:fcdetail}
G^{\Phi\Psi}_{p\sigma,q\sigma} = 
\left(
\begin{array}{cc}
I & 0 \\
0 & G^{\Phi_A \Psi_A} \\
\end{array}
\right)
\end{equation}
where the top-left (bottom-right) block denotes pairs of inactive (active) orbitals.
A simple calculation based on \eqref{eq:fcdetail}, and of similar properties holding for the two-particle
Green's function, shows that the matrix element 
$\frac{\langle\Phi|\hat{H}|\Psi\rangle}{\langle \Phi | \Psi\rangle}$ of $\hat{H}$ is equal to
\begin{equation}
\label{eq:fc_ham}
\frac
{\langle \Phi | \hat{H} |\Psi\rangle}{\langle\Phi| \Psi\rangle}
=
\frac
{\langle \Phi_{A}| H_{A,0}^\prime + \hat{H}_{A,1}^\prime + \hat{H}_{A,2}^\prime | \Psi_{A}\rangle}
{\langle \Phi_{A}| \Psi_{A}\rangle}
\end{equation}
with
\begin{equation}
\begin{split}
H_{A,0}^\prime &= H_{0}+2\sum_{p\in I}h_{pp}  + 
\sum_\gamma 4 \left( l^\gamma \right)^2 - \sum_{ p \in I } 2 \, K^\gamma_{pp}\,, \\
\hat{H}^\prime_{A,1} &= 
\sum_{pq\in A} \Big( h_{pq} + \sum_\gamma 4 l^\gamma L^\gamma_{pq} - 2 K^\gamma_{pq} \Big)
\sum_\sigma \crt{p\sigma} \dst{q\sigma}\,, \\
\hat{H}^\prime_{A,2} &= 
\sum_{ pqrs\in A } \sum_\gamma L^\gamma_{pr} L^\gamma_{qs} 
\sum_{\sigma\tau} \crt{p\sigma} \crt{q\tau} \dst{s\tau} \dst{r\sigma}\,,
\end{split}
\end{equation}
where $l^\gamma = \sum_{p \in I} L^\gamma_{pp}$, 
$K^\gamma_{pq} = \sum_{r \in I} L^\gamma_{pr} L^\gamma_{rq}$. 

\section{Statistical data analysis}

A ph-AFQMC calculation yields a collection of weighted averages
\begin{equation}
w_i = \sum_w W_{i,w} \,\, , \,\, x_i = \sum_w \frac{W_{i,w}}{ \sum_w W_{i,w}} \, \mathcal{E}_{loc}(\Psi_{i,w})
\end{equation}
of the ground-state energy, obtained at simulation times $\beta_i = i \Delta\tau$. 
The latter are outcomes of random variables $X_i$ with mean E$[X_i]$ converging to the 
ground-state energy in the large $i$ limit.

\begin{figure}[ht!]
\centering
\includegraphics[width=0.45\textwidth]{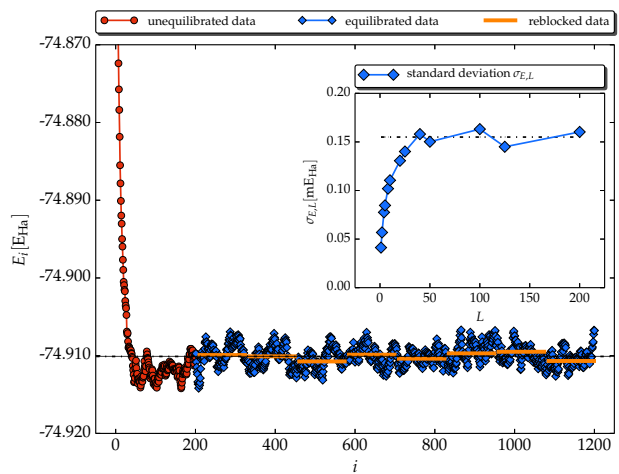}
\caption{(color online) Example of statistical analysis of AFQMC data. After removal of the equilibration phase 
(first 200 samples), data are reblocked until convergence of the standard deviation is reached (shown in the 
inset, and attained for $L \simeq 125$).
}
\label{fig:stat}
\end{figure}

These random variables are neither independent nor identically distributed, as sketched in 
Figure \ref{fig:stat}, so statistical data analysis has to be performed with care.
Unequilibrated data have to be eliminated from the sample, an operation that 
often can be done visually. Equilibrated data are not statistically independent, 
since they are the outcomes of a Markov chain random walk. 

Standard procedures can be applied to characterize the autocorrelation and 
provide a reliable estimate of the statistical error bar.
A way to assess the number $i^*$ of blocks after which data are practically decorrelated
  is to arrange the data $\{x_i,w_i\}_{i=0}^{N-1}$ in 
$N_b$ blocks of length $L$,
\begin{equation}
w^\prime_b = \sum_{i=bL}^{(b+1)L-1} w_i 
\,\, , \,\, 
x^\prime_b =  \frac{ \sum_{i=bL}^{(b+1)L-1} w_i x_i}{\sum_{i=bL}^{(b+1)L-1} w_i}
\end{equation}
and compute the weighted average and variance of the reblocked data,
\begin{equation}
\overline{X} = \frac{\sum_{b=0}^{N_b-1} w^\prime_b \, x^\prime_b}{v_1} \,\, , \,\,
\overline{S}^2 = \frac{ \sum_{b=0}^{N_b-1} w^\prime_b (x^\prime_b - \overline{X})^2 }{ v_1- \frac{v_2}{v_1} } 
\end{equation}
with $v_l = \sum_{b=0}^{N_b-1} \left(w^\prime_b\right)^l$. While the mean $\overline{X}$ is 
independent of $L$, the variance $\overline{S}^2$ saturates towards an asymptotic value 
$\sigma^*$ as $L$ increases. Data reblocked with $L$ sufficiently large are thus independent 
and identically distributed, and $|\mu - \overline{X} | \leq \frac{2\overline{S}}{\sqrt{N_b-1}}$ 
with confidence level 95\%.

%

\end{document}